\documentclass[a4paper,11pt]{article}
\pdfoutput=1 

\usepackage{jheppub}
\usepackage{float}
\usepackage{ulem}
\usepackage[utf8]{inputenc}
\usepackage{mathrsfs}
\usepackage{setspace}
\usepackage{amsfonts}
\usepackage{amssymb}
\usepackage{amsmath}
\usepackage{esint}    
\usepackage{tikz-cd}
\usetikzlibrary{matrix,arrows,positioning,shapes,fadings,decorations.pathmorphing,
decorations.pathreplacing,automata,shadows,decorations.markings,patterns}
\usepackage{epsfig}
\usepackage{latexsym}
\usepackage{color}
\usepackage{graphicx}
\usepackage{hyperref}
\usepackage{nicefrac}
\usepackage{pstricks}
\usepackage{slashed}
\usepackage{multirow}
\usepackage{appendix}
\usepackage{subcaption}

\def\be{\begin{equation}}
\def\ee{\end{equation}}
\def\ba{\begin{eqnarray}}
\def\ea{\end{eqnarray}}

\definecolor{myDeepPurple}{RGB}{95, 5, 120}   
\definecolor{myDeepBlue}{RGB}{15, 35, 110}    
\definecolor{myDeepOrange}{RGB}{170, 60, 0}   
\hypersetup{
    colorlinks = true,
    citecolor  = myDeepPurple,
    linkcolor  = myDeepBlue,
    urlcolor   = myDeepOrange
}

\def\cA{{\cal A}}
\def\cB{{\cal B}}
\def\cC{{\cal C}}

\def\cF{{\cal F}}

\def\IR{\relax{\rm I\kern-.18em R}}

\def\inv{^{\raise.0ex\hbox{${\scriptscriptstyle -}$}\kern-.05em 1}}

\newcommand{\pd}{\partial}
\newcommand{\dr}{\mathrm{d}}

\newcommand{\rmi}{\mathrm{i}}
\newcommand{\Tr}{\text{Tr}}
\newcommand{\tr}{\text{tr}}
\newcommand{\Ad}{\text{Ad}}

\newcommand{\WZWtwo}{\mathrm{WZW}_2}

\newcommand{\bCP}{\mathbb{CP}}
\newcommand{\bPT}{\mathbb{PT}}

\newcommand{\fg}{\mathfrak{g}}
\newcommand{\fh}{\mathfrak{h}}

\def\bb{\ensuremath\mathbb}

\begin{document}

\title{From Diamond Gaugings to Dualisations }

\abstract{~~We revisit the proposal that coupling two six-dimensional holomorphic  Chern–Simons theories generates gaugings throughout the twistor-space diamond relating 6d hCS, 4d self-dual Yang-Mills, 4d Chern–Simons, and 2d integrable models. In previous work this mechanism was demonstrated only in a special case, leaving its general status unclear. By reformulating the construction in the language of Cartan geometry, we expose the underlying gauge structure and show that the argument extends to generic choices of meromorphic data. We then apply this to the pole structure that yields the well-studied $\lambda$-deformations of the WZW model. The coupled 6d system indeed induces gaugings of the associated $\lambda$-models, but necessarily introduces Lagrange multipliers enforcing flatness of the gauged connection. The resulting two-dimensional theories are therefore non-Abelian dualisations rather than ordinary gauged $\lambda$-models. } 

\author{Dimitrios Chatzis, John M. Marley and Daniel C. Thompson}

\affiliation{Centre for Quantum Fields and Gravity, \\ Department of Physics, Swansea University,\\
Swansea SA2 8PP, United Kingdom} 

\emailAdd{d.c.thompson@swansea.ac.uk}
\emailAdd{dchatzis@proton.me}
\emailAdd{2597427@swansea.ac.uk}

\maketitle

\flushbottom
 
\section{Introduction}
Following the proposal by Bittleston and Skinner \cite{Bittleston:2020hfv}, a series of recent developments have demonstrated that six-dimensional holomorphic Chern-Simons, [hCS$_6$], theory on twistor space can serve as a higher-dimensional origin for integrable models.  This theory is described by an action for a $\frak{g}$-valued connection ${\cal A} \in \Omega^{(0,1)} (\mathbb{PT}) \otimes \frak{g}$, 
\begin{equation} \label{hCS6_action_A}
         S_{\mathrm{hCS}_6}[\mathcal{A}]=\frac{1}{2\pi i}\int_{\mathbb{PT}}\Omega\wedge\mathrm{tr}_{\mathfrak{g}}\left(\mathcal{A}\wedge\bar\partial\mathcal{A}+\frac{2}{3}\mathcal{A}\wedge\mathcal{A}\wedge\mathcal{A}\right) \, . 
\end{equation}
 The critical piece of information here is the specification of the $(3,0)$-form $\Omega$.   There is no such global analytic form on projective twistor space\footnote{      
 By contrast, projective supertwistor space, $\mathbb{PT}^{3|4}$, 
 is a super Calabi–Yau manifold admitting a global holomorphic Berezinian. In this setting, (super) holomorphic Chern–Simons theory arises as the open-string field theory of the topological B-model in Witten’s twistor-string construction \cite{Witten:2003nn}, and reproduces the self-dual sector of $\mathcal{N}=4$
super Yang–Mills theory (see also \cite{Sharma:2025ntb,Jarov:2025qhz} for recent progress on the top-down holographic interpretation of this framework).},  $\mathbb{PT}$, hence $\Omega$ is necessarily {\em meromorphic}.   The  specification of the pole structure and boundary conditions on the connection at these poles serves to define different theories.   

By working in a suitable gauge and eliminating components constrained by the equations of motion, the hCS$_6$ action localises to the poles of $\Omega$. This produces a four-dimensional theory whose equations of motion can be recast as those of self-dual Yang–Mills theory.  This provides a Lagrangian implementation of the Ward–Penrose correspondence \cite{Ward:1977ta}, which relates holomorphic bundles on twistor space to self-dual connections in four dimensions.   
Further dimensional reduction yields a two-dimensional theory, in which the Lax connection descends naturally from the original six-dimensional gauge field. The precise nature of the resulting integrable model depends on the six-dimensional data (poles and boundary conditions) and on the choice of reduction directions, along which fields are taken to be invariant.  For example, when $\Omega$ has two double poles, the resulting theory is the principal chiral model with a Wess–Zumino term, whose level is determined by the reduction data \cite{Bittleston:2020hfv}. Splitting one of the double poles into a pair of single poles generates  \cite{Cole:2023umd}  a well-studied class of integrable deformations of the WZW model known as the 
$\lambda$-model of Sfetsos \cite{Sfetsos:2013wia}.   

Alternatively, one can first reduce to the four-dimensional Chern-Simons theory of Costello, Witten, and Yamazaki \cite{Costello:2017dso,Costello:2018gyb,Costello:2019tri}, and then apply a similar localisation procedure. Both paths lead to the same two-dimensional integrable model, yielding a commutative “diamond” of theories

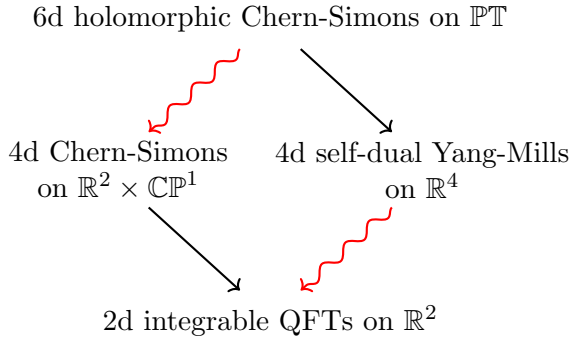
\begin{figure}[h!]
\centering
\begin{tikzpicture}
\node at (0,2) { $\text{6d holomorphic Chern-Simons on}\,\, \mathbb{PT}$};
\node[align=center] at (-2,0) { 4d Chern-Simons \\ on $\mathbb{R}^2\times\mathbb{CP}^1$}; 
\node[align=center] at (2,0) { 4d self-dual Yang-Mills\\ on $\mathbb{R}^4$};

\node[align=center] at (0,-2) { 2d integrable QFTs on $\mathbb{R}^2$};
\draw[->,thick,red,decorate, decoration={snake, segment length=12pt, amplitude=2pt}] (-0.4,1.6)--(-1.6,0.5);
\draw[->,thick] (0.4,1.6)--(1.6,0.5);
\draw[->,thick,red,decorate, decoration={snake, segment length=12pt, amplitude=2pt}] (1.6,-0.5)--(0.4,-1.6);
\draw[->,thick] (-1.6,-0.5)--(-0.4,-1.6);
\end{tikzpicture}
\caption{The diamond of theories spawned by  \eqref{hCS6_action_A}. The black straight arrows denote localization on $\mathbb{CP}^1$ while the red wavy ones dimensional reduction along certain directions in $\mathbb{R}^4$.}
\label{diagram:diamond}
\end{figure}

Whilst this diamond correspondence has been fully fleshed out for a number of examples there are also a variety of integrable models for which some elaborations will be required.   Notably,  many integrable non-linear sigma models arise via gauging—for instance, gauged WZW models, which realise coset CFTs, and their integrable $\lambda$-deformations.  In recent work  \cite{Cole:2024sje}, it was shown that the diamond construction could itself be extended to incorporate gauging. As we shall review in section \ref{section:gauged_hCS}, the motivation for this construction lies in the way a $G/H$ gauged WZW model can be obtained, via the Polyakov-Wiegmann identity  \cite{Polyakov:1984et}, as the difference of a $G$-WZW and   $H$-WZW.   Thus \cite{Cole:2024sje} (and see \cite{Stedman:2021wrw} for related considerations in CS$_4$) one considers a 6d action containing additionally an $\frak{h}$-valued connection ${\cal B} \in \Omega^{(0,1)} (\mathbb{PT}) \otimes \frak{h} $ of the form

\begin{equation}
  S_{\mathrm{ghCS_6}}[{\cal A},{\cal B}]  = S_{\mathrm{hCS_6}}[{\cal A} ] - S_{\mathrm{hCS_6}}[{\cal B} ] + S_{\text{int}}[{\cal A},{\cal B}]  \, ,  \nonumber
\end{equation}
where there is a ``boundary'' (i.e. localised to poles of $\Omega$) interaction\footnote{Left implicit in these equations  is a choice of homomorphism $\rho:\mathfrak{h}\to\mathfrak{g}$ defining the embedding of the sub-algebra.  In this work we consider uniquely vectorial gaugings; in \cite{Cole:2024sje} other gaugings were reached by allowing  $\rho$ to be defined piecewise at each pole.   }
\begin{equation}\label{interaction_term}
S_{\text{int}}[{\cal A},{\cal B}] = -\frac{1}{2\pi i} \int \bar\partial \Omega \wedge \mathrm{tr}\left(\mathcal{A}\wedge \mathcal{B} \right) \, .\nonumber
\end{equation}
 The theory has to be completed by the specification of boundary conditions that ensure the vanishing of the boundary term produced upon varying the action. 
 
The calculation then proceeds by  introducing edge modes, localised to the poles of $\Omega$,  to compensate for otherwise broken symmetries.   In this way, the entire action can be localised to contributions at these poles in terms of the edge modes which source the dynamical degrees of freedom. In \cite{Cole:2024sje} this idea was elaborated and studied for examples when $\Omega$ has two double poles;  even in this scenario the interplay between reduction data and constraints within the system gave rise to quite a rich tapestry of integrable models based on coset CFTs.

The aim of the present work is to refine and simplify the construction of gauged integrable models within the six-dimensional holomorphic Chern--Simons framework. While the approach of \cite{Cole:2024sje} provides a valuable proof of concept, it suffers from several limitations: the boundary interaction term appears \textit{ad hoc}, the emergence of the gauged degrees of freedom relies on intricate cancellations, and the relationship between boundary conditions and the solutions of the gauged and ungauged models remains opaque. As a consequence, explicit analysis has so far been restricted to the simplest meromorphic data, namely $\Omega$ with two double poles. 

We overcome these limitations by presenting a significantly simpler and more systematic formulation of the gauged diamond correspondence. Our construction clarifies the role of boundary terms, identifies the gauged degrees of freedom directly, and makes gauge covariance and the propagation of boundary conditions manifest. This allows us to treat a broader class of meromorphic forms and associated integrable models in a conceptually transparent and technically tractable manner. 
The central technical insight is  a reframing of the fundamental variables:  we decompose the six-dimensional connection as ${\cal A} = {\cal B} + {\cal C}$, where ${\cal B}$ corresponds to an unbroken gauge symmetry.  ${\cal C}$ is a Cartan-type connection, transforming tensorially in the adjoint representation under the local $\mathfrak{h}$ symmetry for which ${\cal B}$ is the connection. This decomposition makes gauge covariance manifest and eliminates the need for auxiliary edge modes that can be fixed away. Interactions between ${\cal C}$ and ${\cal B}$ ensure that upon reduction, flatness constraints are automatically enforced in the lower-dimensional dynamics, which in two dimensions corresponds to gauging with a flat connection, i.e., a Buscher-type dualisation \cite{Buscher:1987sk}.  This is a central result of the present work: \textit{the gauged diamond implements a dualisation}.   

In this framework, the boundary term introduced in \cite{Cole:2024sje} is no longer \textit{ad hoc}: it naturally arises to cancel a total derivative that would otherwise appear, making the construction manifestly gauge covariant. This clarifies the underlying structure.  
 
As a concrete application, we consider the pole structure of $\Omega$ previously shown to yield $\lambda$-deformed WZW$_2$ models \cite{Cole:2023umd}, following the right-hand side of the diamond (Figure \ref{diagram:diamond}). In the ${\cal B}/{\cal C}$ decomposition, the resulting two-dimensional theory is expressed in gauge-covariant variables. It corresponds to the  non-Abelian T-dual of the two-field $\lambda$-model, with the additional degrees of freedom in earlier formulations now understood as components of the Cartan-type connection ${\cal C}$ enforcing flatness.  At a specific choice of parameters, the theory acquires an additional local invariance, allowing one of the two edge mode fields to be eliminated. The resulting model is the dual of the standard $\lambda$-model. A Lax formulation is obtained directly from the dimensional reduction of the higher-dimensional gauge connections. The integrability of the dual theory is guaranteed at the classical level considered here, as it follows from the canonical nature of the underlying duality transformation.  We also clarify a subtle but important point. Although one may relax the flatness constraint--leading to a well-defined gauging of the integrable $\lambda$-deformed WZW model--the resulting theory is generically distinct from the integrable $\lambda$-deformation of the $G/H$ gauged WZW model.
We illustrate these results explicitly with an example based on the $\mathrm{SU}(2)/\mathrm{U}(1)$ coset theory.

\section{Cartan Connections and Chern-Simons}\label{background_field_variables}

Instead of working with the connections ${\cal A} $ and ${\cal B}$ we will consider the expansion of ${\cal A}$ around ${\cal B}$ by introducing ${\cal C} = \cA - \cB$.
This is exactly akin to how one might expand ${\cal A}$ around some fixed background, but here of course ${\cal B}$ is a field that will be integrated over. In what follows the fact that these connections are holomophic is not vital, so we will make general statements that can be applied to the hCS context with trivial changes restricting fields and exterior derivatives to holomorphic counter parts.   Whilst the ingredients of this section are, of course, somewhat standard we have not found a reference precisely aligned to our requirements so we synthesise these ideas below without claiming originality. 

There are two local symmetries,  the first is a local $H$-symmetry under which $\cB$ transforms as a connection one-form, but for which ${\cal C}$ transforms adjointly (the derivative of gauge parameter cancelling between contributions from ${\cal A}$ and ${\cal B}$) 
\begin{align}\label{eq:Hsymm}
    h: \cB \mapsto    \cB^{(h)} \equiv    h^{-1} \dr h + h^{-1} \cB h \, , \quad {\cal C} \mapsto h^{-1} {\cal C} h\, .  
\end{align}
The second is a $G$-symmetry under which $\cB$ is invariant but for which 
\begin{equation} \label{eq:Gsymm}
   g:  {\cal C} \mapsto g^{-1}\nabla g +  g^{-1} {\cal C} g \, ,
\end{equation}
where $\nabla \bullet   = \mathrm{d}  \bullet + [\cB,\bullet]  $. A key feature however is that the varied ${\cal C}$ depends on ${\cal B}$; such that the composition of $H$ and $G$ symmetries becomes non-trivial.  It is natural to invoke that the $ g\in G$ gauge parameters transform under the $H$ action as 

\begin{equation}\label{eq:gcompensator}
    h: g \mapsto h^{-1} g h \, ,
\end{equation} 
such that  $g^{-1} \nabla g \mapsto  h^{-1} (g^{-1} \nabla g ) h $ and the transformations thus compose to a commutative diagram:
\[
\begin{tikzcd}[row sep=3em, column sep=4em]
{\cal C} 
  \arrow[r, "g"] 
  \arrow[d, "h"'] 
& g^{-1} \nabla g + g^{-1} {\cal C} g 
  \arrow[d, "h"'] \\
h^{-1} {\cal C} h 
  \arrow[r, "g"'] 
& h^{-1}  \left(g^{-1} \nabla g + g^{-1} {\cal C} g   \right) h
\end{tikzcd}
\]
Applied to the Chern-Simons three-form, $ {\cal L}_{\text{CS}}$, we have the crucial identity 
\begin{align}\label{Ltotal}
{\cal L}_{\text{tot}} [{\cal A},{\cal B}] & \equiv  {\cal L}_{\text{CS}}[{\cal A}]  -  {\cal L}_{\text{CS}}[{\cal B}] - \dr \,  \mathrm{tr}(\cA \wedge \cB ) 
\nonumber\\ & = {\cal L}^\nabla_{\text{CS}}[{\cal C}]  + 2  \mathrm{tr}({\cal C}  \wedge {\cal F}[\cB] )  \equiv {\cal L}_{\text{gCS}}[{\cal C} ,{\cal B}] \, ,
\end{align}
in which $ {\cal L}^\nabla_{\text{CS}}$ is the Chern-Simons three form with the replacement of derivative to covariant derivative i.e. 
\begin{equation} \label{LCS}
{\cal L}^\nabla_{\text{CS}}[{\cal C},{\cal B}] =  \mathrm{tr} \left({\cal C}\wedge\nabla {\cal C} + \frac{2}{3} {\cal C} \wedge  {\cal C} \wedge  {\cal C}  \right) \, .  
\end{equation}
Notice from this perspective the inclusion of the interaction term in ${\cal L}_{\text{tot}} $ is essential to avoid total derivatives after recasting in terms of ${\cal C}$. 

It is immediate to see, because $\nabla {\cal C}$ is an appropriate covariantised derivative, that under the $H$-symmetry of eq.~\eqref{eq:Hsymm} the Lagrangian \eqref{Ltotal} is absolutely invariant, $\delta_h {\cal L}_{\text{gCS}}[{\cal C} ,{\cal B}] = 0$, without any boundary terms generated. This is crucial when we consider the holomoprhic Chern-Simons theory; as any total derivative contributions to gauge variations source edge modes, the absence of such for the $H$-symmetry means that there is no reason to invoke a Stueckelberg edge mode prescription.  

The transformation under the $G$-symmetry of eq.~\eqref{eq:Gsymm} is more delicate, 
\begin{equation} 
\delta_g {\cal L}_{\text{gCS}}[{\cal C} ,{\cal B}] =  \dr  \mathrm{tr} \left({\cal C} \wedge  \nabla g g^{-1}  \right) - {\cal L}^{(3)}_{\text{gWZ}}[g, \cB] \, , \nonumber
\end{equation} 
in which we define the three-form 
\begin{equation}\label{LgWZ}
{\cal L}^{(3)}_{\text{gWZ}}[g, \cB] =  \mathrm{tr}\left[  \frac{1}{3} \left( g^{-1} \dr g  \right) ^3   + \dr  \left( g^{-1} \dr g \wedge \cB + \dr g g^{-1}\wedge \cB +  \cB  \wedge g \cB  g^{-1}  \right)\right] \, . \nonumber
\end{equation}
Here again we see the point of the compensating transformation eq.~\eqref{eq:gcompensator} of $g$ under the $H$ symmetry since $ {\cal L}^{(3)}_{\text{gWZ}}[g, \cB]$ is the gauge invariant completion of the WZ three-form:
 $$
 {\cal L}^{(3)}_{\text{gWZ}}[ h^{-1}g h , \cB^{(h)}] =    {\cal L}^{(3)}_{\text{gWZ}}[  g , \cB ] \, . 
 $$
The anomalous terms under the $G$ action echo precisely those produced when a standard Chern-Simons three-form undergoes gauge transformation, but now appear in a $H$-symmetry invariant fashion.  Since we already demonstrated that $\delta_h  {\cal L}_{\text{gCS}}[{\cal C} ,{\cal B}] =0 $   performing a subsequent transformation $\delta_g \delta_h {\cal L}_{\text{gCS}}[{\cal C} ,{\cal B}]  =0  $, and the commutative diagram ensures that  $\delta_h \delta_g {\cal L}_{\text{gCS}}[{\cal C} ,{\cal B}]  =0 $.

The above can be phrased in a more mathematically direct fashion.  We begin with two principal bundles $G\hookrightarrow P_G \twoheadrightarrow M $ and $H\hookrightarrow P_H \twoheadrightarrow M $  equipped with connections  for which  ${\cal A}$ and  ${\cal B}$ are the local connection one-forms  i.e. the gauge fields.  We assume  a homomorphism $\rho:\mathfrak{h}\to \mathfrak{g}$ which lifts in an obvious way to define an embedding of bundles $P_H \subset P_G$.  The field ${\cal C}$ transforms as a section of the associated adjoint bundle; we form the associated product bundle $P_H \times_H \frak{g}$ by imposing an equivalence on the direct product $P_H \times \frak{g}$ that $(p \triangleright h , X) \sim ( p , \Ad_h^{-1} X)  $ for all points.  This bundle over $M$ then has fibres that are $\frak{g}$ but are patched using the $H$-valued transition functions.  Whilst there is an inclusion $P_H \times \frak{g}\hookrightarrow \Ad P_G$ it is important to emphasize that it is the transition functions are the ones of the reduced structure group.     This then provides a distinguished splitting of connection on $P_G$ relative to the reduction $P_H \subset P_G$ under which ${\cal A} = {\cal B} + {\cal C}$ for which ${\cal C}$ is tensorial and ${\cal B}$ an $H$-connection.    

The inclusion of the total derivative term in ${\cal L}_{\text{tot}}[{\cal A}, {\cal B}]$ also can be given an informative rationale here.  Turning to the characteristic class interpretation \cite{ChernSimons:1974}, recall that the Chern–Simons three-form ${\cal L}_{\text{CS}}[\cA]$ is a secondary characteristic class defined such that its derivative gives the second Chern class  $\dr {\cal L}_{\text{CS}}[\cA] = \tr {\cal F}[\cA] \wedge  {\cal F}[\cA] $.   In general the Chern-Weil theorem shows the difference between second Chern classes of two connections is exact, and is encoded in a transgression 
$$
\dr Q^{(3)}[\cA, \cB] = \tr {\cal F}[\cA] \wedge  {\cal F}[\cA] - \tr {\cal F}[\rho(\cB)] \wedge  {\cal F}[\rho(\cB)]  \, .
$$
A canonical presentation of the transgression $n$-th Chern classes is given by\footnote{We adapt Lemma 5, p.~297 of \cite{KobayashiNomizu:1969}, which constructs the transgression form for the straight-line interpolation between two connections. Since the space of connections is affine, one can take the origin at zero to make direct contact with the canonical Chern--Simons forms of \cite{ChernSimons:1974}.}
$$
 Q^{(2n-1)} [\cA, \cB] =n  \int_0^1 \mathrm{d}t\,  \tr \left(  {\cal C} \wedge {\cal F}[\rho(\cB) +  t {\cal C}]^{n-1} \right) \, .
$$
Specialising to the case of $n=2$ and evaluating the integral one finds immediately that ${\cal L}_{\text{tot}}[{\cal A}, {\cal B}]$ is precisely this canoncial transgression form. In fact, our explicit computation of the finite gauge variation of ${\cal L}_{\text{tot}}[{\cal A}, {\cal B}]$  realises the standard descent procedure: the variation decomposes into an exact form together with the canonical Wess–Zumino cocycle. In this sense our formula completes the descent construction for the transgression form relevant here (in the sense of the finite gauge variation).

\section{Ungauged holomorphic Chern-Simons theory on twistor space}
Briefly let us review the salient features of the ungauged holomorphic Chern-Simons theory on twistor space $\mathbb{PT}$.  This is a theory defined by a connection 
$  
    \cA \in \Omega^{(0,1)}(\bPT)\otimes \fg \, ,  
$   
and a meromorphic  $(3,0)$-form $\Omega\in \Omega^{(3,0)}(\bPT)$ with an action given by \eqref{hCS6_action_A}. This theory forms the highest dimensional node of the diamond depicted in figure \ref{diagram:diamond}. Truncation of \eqref{hCS6_action_A} to the zero modes of a complex vector field results in a reduction of the theory to the well studied 4d Chern-Simons theory which in turn is equivalent (by a ``localisation" process analogous to the one we shall shortly present) to integrable (and potentially conformal) two-dimensional quantum field theories, the details of which depend on the both the pole structure of the starting $\Omega$ and the reduction data\footnote{From the 4d perspective $\Omega$ descends to a meromorphic differential on a curve with both poles and zeros according the Riemann Roch theorem.  The location of poles and zeros, and the order and residues at the poles specify the 2d theory.}. This procedure outlines the left hand side of the diamond, about which we shall say no more in this work.

Our predominant interest is in the right hand side of the diamond.  Here we first obtain a four-dimensional integrable theory (a self-dual Yang-Mills) localised to the poles of $\Omega$.   To proceed along this route we need a few technical conventions.   For this we consider\footnote{To be more precise, we define $\mathbb{PT}= \mathbb{CP}^3 \setminus \mathbb{CP}_\infty^1$.  Embedding $\mathbb{CP}^3\subset \mathbb{C}^4$  we have holomorphic rays $Z^A = (\pi^a \, , \omega_{\dot{a}})$  described by two spinors subject to the incidence relation $x^{a \dot{a}}  \pi_a = \omega^{\dot{a}}$.   Although $x_{a \dot{a}}$ in general define complexified space-time, we  impose reality conditions by restricting to a slice invariant under a quartic involution $Z^A\rightarrow \hat{Z}^A = ( \hat{\pi}_a , \hat{\omega}^{\dot{a}} )$ such that $x^{a \dot{a}}$ define an affine chart over $\mathbb{R}^4$.
Removing $\mathbb{CP}_\infty^1$, (deletion of the line over the compactification point $x\rightarrow \infty $ corresponding to the $\pi^a = 0 $ spinor), ensures the fibration over  $\mathbb{R}^4$ is well defined.  }   $\mathbb{PT}$ as a $\mathbb{CP}^1$ bundle over a four-dimensional space $\mathbb{E}$.  We invoke homogenous coordinates $\pi^a$ on  $\mathbb{CP}^1$  and coordinates on  $\mathbb{E}$ are 
$$
x^{a \dot{a}} =\frac{1}{\sqrt{2}} \begin{pmatrix}
    x_0 + i x_1  & x_2 + i x_3 \\ -x_2 + i x_3  & x_0 - i x_1  
\end{pmatrix}\, . 
$$
We work in a basis of holomorphic $(1,0)$ and $(0,1)$ forms 
\begin{equation}
\begin{aligned}
& e^0 = \langle \pi \dr \pi \rangle ~, &&
& e^{\dot{a}} = \pi_{a} \dr x^{a \dot{a} } ~, \quad 
& \bar{e}^0 = \frac{\langle \hat{\pi} \dr \hat{\pi} \rangle}{\langle \pi \hat{\pi} \rangle^2} ~, &&
& \bar{e}^{\dot{a}} = \frac{\hat{\pi}_{a} \dr x^{a \dot{a}   }}{\langle \pi \hat{\pi} \rangle} \,, \nonumber
\end{aligned}
\end{equation}
 in which we denote the conjugated spinor   $\hat{\pi}_a = ( -\bar{\pi}_2 , \bar{\pi}_1)$  and the contraction $\langle \pi \chi \rangle = \epsilon^{ab} \pi_b \chi_{a} = \pi^a \chi_{a} = \pi_1 \chi_2 - \pi_2 \chi_1 $.  The differential we consider has the form $\Omega = \phi(\pi ) e^0 \wedge e^{\dot{1}} \wedge  e^{\dot{2}}$
 and is defined by a homogeneous function. To ensure that we have a well defined object on $\mathbb{CP}^3$ we require that $\Omega$ be invariant under the relation $\pi^a \sim \lambda \pi^a$, and as the holomorphic forms carry weight,  we require that $\phi(\pi) $ be degree minus $4$ and so $\Omega$ will necessarily have poles in   $\mathbb{CP}^1$.     Informative examples are the choices of either two double poles, relevant for undeformed WZW CFT$_2$, 
 $$
\textrm{WZW-type:} \quad   \phi = \phi_{2,2} = \frac{K}{\langle  \alpha \pi \rangle^2 \langle  \beta\pi  \rangle^2} \, ,
 $$
 or splitting one of the double poles into two single poles, relevant for $\lambda$-deformation of WZW CFT$_2$,   
 \begin{equation}\label{112_pole_structure}
\textrm{$\lambda$-type:} \quad   \phi = \phi_{1,1,2} = \frac{K}{\langle  \alpha \pi \rangle \langle \tilde{\alpha} \pi \rangle \langle  \beta\pi  \rangle^2} \, ,
 \end{equation}
In both examples the critical point here is that $\bar{\pd} \Omega$ will no longer vanish but instead have some support on the poles.  This means that although the bulk equations of motion are $\Omega \wedge {\cal F}[{\cal A} ] = 0 $, boundary conditions are required to enforce the vanishing of the boundary term, 
 $$
    \int _{\bPT} \bar{\pd} \Omega \wedge  \tr \left( \cA \wedge \delta \cA  \right) \, ,   
$$ 
 obtained from varying the action.  The boundary conditions can be solved locally at each pole, e.g. in the WZW-type imposing 
 $$
 {\cal A} \vert_{\pi = \alpha, \beta} = \pd_0   {\cal A} \vert_{\pi =  \alpha, \beta} =0 \, ,
 $$
or be more intricate ensuring cancelling contributions across poles.  This more involved type of condition occurs in the $\lambda$-type pole structure whereby one can impose 
$$
  {\cal A} \vert_{\pi = \alpha} = {\cal A} \vert_{\pi = \tilde{\alpha}} \, . 
 $$
 In fact, in the later case it was shown in \cite{Cole:2023umd} that there are even more intricate boundary conditions where different components of the gauge field are related between the two poles with extra scaling freedoms, and these allow for the construction of  which we shall detail in an example later. 

We now consider the would-be gauge symmetry  $\hat{f}\in G $
$$
\hat{f}: \quad {\cal A} \mapsto   \hat{f}^{-1} \bar{\pd} \hat{f} +  \hat{f}^{-1} {\cal A}  \hat{f} \, , 
$$
in which the hats are to remind the reader that these are local over the entirety of $\mathbb{PT}$.   The action is not strictly invariant, instead picks up a boundary term 
\begin{align}
    \delta_{\hat{f}}    S_{\mathrm{hCS_6}}&=  \frac{1}{2\pi i}\int _{\bPT} \bar{\partial} \Omega\wedge      \mathrm{tr} \left({\cal A} \wedge  \bar{\partial } \hat{f} \hat{f}^{-1}  \right) - \Omega \wedge {\cal L}_{\text{WZ}}[\hat{f} ]  \, .  \nonumber
\end{align}

The presence of a ``boundary" changes the nature of these symmetries.  The true gauge redundancies are transformations that preserve the boundary conditions.  Transformations on the other hand that do not preserve these boundary conditions seed global symmetries of dynamical edge modes located at the boundary.  The most familiar example of this ``localisation" is of course 3d CS on a boundary whose edge modes describe (chiral)  WZW theories with an associated affine symmetry.  

The next step then is to extract the relevant edge modes, and this is readily done using a Stueckelberg trick.  By upgrading a gauge parameter to a   field one obtains a parameterisation of ${\cal A}$ in terms of a new ${\cal A}'$ and $\hat{g}\in C^{\infty}(\mathbb{PT},G)$
$$
{\cal A} \equiv \cA^{\prime \hat{g} }=   \hat{g}^{-1} \bar{\pd} \hat{g} +  \hat{g}^{-1} {\cal A}'  \hat{g}\, . 
$$
Now the original would-be gauge symmetry leaves ${\cal A}'$ invariant but acts as 
$$
\hat{f}: {\cal A}' \mapsto   {\cal A}' \, , \quad   \hat{g}  \mapsto  \hat{g}  \hat{f} \, . 
$$
This parameterisation introduces a new redundancy (called {\em internal}-gauge symmetry in \cite{Cole:2023umd}) under which the original ${\cal A}$ is invariant but 
$$
\check{f}: {\cal A}' \mapsto      \check{f}^{-1} \bar{\pd}  \check{f} +   \check{f}^{-1} {\cal A}'   \check{f}  \, , \quad \hat{g}  \mapsto  \check{f}^{-1} \hat{g}\, .  
$$
This internal redundancy is partially fixed by demanding that  the new ${\cal A}'$ field has no-legs on $\mathbb{CP}^1$, ${\cal A}'_0 =0 $. We also typically fix the field $\hat{g}$ to be the identity at one double pole.  The values of $\hat{g}$, and potentially its $\mathbb{CP
}^1$ derivatives, at the remaining poles constitute a set of edge modes.  One then finds that the remaining symmetries, i.e. those compatible with fixing and boundary conditions,  act on $\hat{g}$ as either global or semi-local (with  constrained coordinate dependence) symmetries that act on these edge mode degrees of freedom.   To extract the edge mode dependence we utilise the standard result for the gauge transformation of a Chern-Simons term,
\begin{equation}
  S_{\text{hCS}_6}[\cA] =      S_{\text{hCS}_6}[\cA^{\prime \hat{g} }] =   S_{\text{hCS}_6}[\cA^{\prime  }] + \frac{1}{2\pi i } \int_\mathbb{PT} \Omega \wedge \tr \left(  \bar{\pd } ( {\cal A}' \wedge \bar\partial \hat{g} \hat{g}^{-1} ) - \frac{1}{3} (\hat{g}^{-1}\bar{\pd } \hat{g})^3  \right) \nonumber \, .
\end{equation}
After integration by parts one immediately sees that the final two terms will localise to contributions from the poles involving the edge-modes.  

The procedure then continues by noting that away from the location of the poles the equations of motions enforce that  the components  ${\cal A}'_{\dot{a}} $ in ${\cal A}' ={\cal A}'_{\dot{a}} \bar{e}^{\dot{a} }  $ be holomorphic, and since they must have scaling weight $+1$ they can be expressed in terms of the components of a 4d gauge field as ${\cal A}'_{\dot{a}} = \pi^a A_{a \dot{a}}(x) $.  With this in mind it is then possible to solve the boundary conditions to algebraically determine   $ A_{a \dot{a}}(x) $  in terms of the edge mode at the poles.

\section{Gauged holomorphic Chern-Simons theory on twistor space}\label{section:gauged_hCS}

There is a wide variety of integrable models in two dimensions that can be thought of as having gauge symmetry (not only e.g. principal chiral theories on geometric cosets, but also gauged WZW, Pohlmeyer reduced theories, lambda deformations).  One should anticipate that they can also be seeded within the six-dimensional perspective.  

In \cite{Cole:2024sje} this was approached from a bottom up perspective, starting with a gauged WZW model in two dimensions.  These enjoy a remarkable property; they can be expressed as a difference of two ungauged WZW terms. This is captured via the Polyakov-Wiegmann identity \cite{Polyakov:1984et} 
\begin{equation}
    S_{\text{g}\WZWtwo}[g,B]=S_{\WZWtwo}[\tilde g]-S_{\WZWtwo}[\tilde h] \, ,\nonumber
\end{equation}
where the various fields take values in: $g\in C^{\infty}(\Sigma,G)$, $B\in\Omega^1(\Sigma)\otimes\fh$, $\tilde g\in C^{\infty}(\Sigma,G)$, $\tilde h\in C^{\infty}(\Sigma,H)$, and the WZW action is 
\begin{equation}
    S_{\text{WZW}_2}[g]=\frac{1}{2}\int _{\Sigma}\tr\, \left( g^{-1}\mathrm{d}g\wedge\star  g^{-1}\mathrm{d}g\right) + \frac{1}{3}\int_{{\cal M}_3} \tr \left( \, \hat g^{-1}\mathrm{d}\hat g\wedge \hat g^{-1}\mathrm{d}\hat g\wedge\hat g ^{-1}\mathrm{d}\hat g  \right) \, .\nonumber
\end{equation}
with $\partial {\cal M}_3 = \Sigma$ and $\hat{g}$ an appropriate extension. 
The same concept is generalized, although some details enter the picture, to four-dimensional gauged WZW models \cite{Losev:1995cr}, where the WZW action takes the form:
\begin{equation}
    S_{\text{WZW}_4}[g] = \frac{1}{2}\int_{\mathbb{R}^4} \tr\, \left( g^{-1}\mathrm{d}g\wedge\star g^{-1}\mathrm{d}g \right) + \int_{\mathbb{R}^4\times [0,1]} \omega\wedge\tr \,( \hat g^{-1}\mathrm{d}\hat g\wedge \hat g^{-1}\mathrm{d}\hat g\wedge\hat g ^{-1}\mathrm{d}\hat g) \,,\nonumber
\end{equation}
with $\omega$ being a K\"{a}hler form on $\mathbb{R}^4$. In this case the four-dimensional analogue of the Polyakov-Wiegmann identity holds only when the gauge connection is flat.

This {\em difference of theories} structure motivated the approach of \cite{Cole:2024sje} wherein the difference of two holomorphic Chern-Simons terms was proposed as the avatar of such integrable models.   The  fundamental fields of the theory are now  the connections
\begin{equation}
    \cA \in \Omega^{(0,1)}(\bPT)\otimes \fg \, ,\quad \cB \in \Omega^{(0,1)}(\bPT)\otimes\fh \, , \nonumber
\end{equation}
which are combined in the action 
\begin{equation}\label{ghCS6_action_AB}
     S_{\text{ghCS}}[\cA,\cB]=\frac{1}{2\pi i} \int _{\bPT}\Omega \wedge \Big({\cal L}_{\text{CS}} [\cA] -{\cal L}_{\text{CS}} [\cB]\Big)+ S_{\text{int}}[\cA,\cB] \, ,\nonumber
\end{equation}
where we have the  \textit{ad hoc} introduction of the interaction term  
\begin{equation}
\begin{split}
&S_{\text{int}}[\cA,\cB]=-\frac{1}{2\pi i}\int _{\bPT}\bar\partial\Omega\wedge\tr\left( \cA\wedge \cB \right) \, .
\end{split}\nonumber
\end{equation}
We now follow broadly the procedure laid out in the ungauged model.   The first step is to  note the boundary terms coming from variation of the action 
\begin{equation}\label{variation_CS_A_B}
\begin{split}
\int_{\mathbb{PT}}\bar\partial\Omega\wedge\mathrm{tr}\left[\left(\delta\mathcal{A}+\delta\mathcal{B}\right)\wedge\left(\mathcal{A}-\mathcal{B}\right) \right] \, . 
\end{split}\nonumber
\end{equation}
Here one encounters a puzzle.   Whilst it is clear that the WZW-type boundary condition of the ungauged theory can be readily lifted to this term, it appears at first sight rather more challenging to apply the more general cases of boundary conditions that relate different poles.    However it was seen in \cite{Cole:2024sje} that for the case of two double poles that we can solve the boundary condition in terms of edge modes of ${\cal A}$, but can't determine ${\cal B}$;  indeed a careful study of the gauge symmetries shows that the ${\cal B}$ sources a four-dimensional gauged field that covariantises derivatives of the edge mode with respect to the $H$-local symmetry.  All steps involved required significant amount of algebra especially since symmetries are not made manifest during the calculation. 

Instead of this rather cumbersome procedure we will utilise the background field variables of section \ref{background_field_variables}, namely we shall work with ${\cal B}$ and the combination 
\begin{equation}
    {\cal C } = {\cal A} -  {\cal B}  \, . \nonumber
\end{equation}
This provides an immediate justification for the inclusion of the interaction term $S_{\text{int}}$ since it eliminates the production of boundary terms in when changing variables such that 
\begin{equation}
       S_{\text{ghCS}}[\cA,\cB]=  \frac{1}{2\pi i}\int _{\bPT} \Omega\wedge {\cal L}_{\text{CS}}^{\nabla}[\cC,\cB]+ \frac{1}{\pi i}\int _{\bPT} \Omega\wedge\tr\, \left( \cC\wedge \cF[\cB] \right) \, , \nonumber
\end{equation}
where we invoke the gauged Chern-Simons three-form defined in \eqref{LCS}.
This presentation makes manifest local $H$-gauge transformations parametrised by elements  $\hat{h}\in C^{\infty}(\bPT,H)$ which act as 
\begin{equation}
  \hat{h}:  \quad  \cB\mapsto  \hat{h}^{-1}\bar\partial  \hat{h}+\mathrm{Ad}_{\hat{h}}^{-1}\cB,\quad \cC\mapsto  \mathrm{Ad}_{ \hat{h}}^{-1}\cC, \nonumber
\end{equation}
and under which   the anti-holomorphic covariant derivative, $\bar{\nabla} = \bar{\partial} + \cB$ ,  of $\cC$ transforms adjointly, $\bar{\nabla}\cC\mapsto  \mathrm{Ad}_{ \hat{h} }^{-1}(\bar{\nabla}\cC)$.
Since this transformation holds   is exact and does not generate boundary terms,  it does not produce edge modes and there is no need from the outset to perform any Stueckelberg type analysis on the $\cB$ gauge field. 

In contrast however under the $G$-transformations parametrised by $\hat{f}\in C^{\infty}(\bPT,G)$ 
\begin{equation}
   \hat{f}: \quad   \cC\to      \hat{f}^{-1}\bar{\nabla}   \hat{f} + \Ad_{   \hat{f}}^{-1}\cC \, , \nonumber
\end{equation}
do not lead the action strictly invariant but instead produce the contributions 
\begin{align*}
    \delta_{\hat{f}}    S_{\text{ghCS}}&=  \frac{1}{2\pi i}\int _{\bPT} \bar{\partial} \Omega\wedge      \mathrm{tr} \left({\cal C} \wedge  \bar{\nabla} \hat{f} \hat{f}^{-1}  \right) -  \Omega \wedge {\cal L}^{(3)}_{\text{gWZ}}[\hat{f}, \cB]   \\ 
    &= \frac{1}{2\pi i}\int _{\bPT} \bar{\partial} \Omega\wedge      \mathrm{tr}  \left({\cal C} \wedge  \bar{\nabla} \hat{f} \hat{f}^{-1}  - \hat{f}^{-1}  \bar{\partial}  \hat{f} \wedge \cB- \bar{\partial} \hat{f} \hat{f}^{-1}\wedge \cB-  \cB  \wedge \hat{f} \cB  \hat{f}^{-1}  \right)  -\Omega \wedge {\cal L}_{\text{WZ}}[\hat{f}] \, .
\end{align*}   
We see that the   ${\cal B}$ enters here to simply $H$-covariantise the anomalous terms that were already present in the ungauged theory.   So our consideration of the ${\cal C}$ field will proceed, \textit{mutatis mutandis}, in the same fashion as the treatment of ${\cal A}$ in the ungauged theory with the following steps
\begin{enumerate}
    \item {\em Stueckelberg fields} To   restore exact invariance under 
G-transformations  we will introduce a set of Stueckelberg modes $\hat{g}\in C^{\infty}(\bPT,G)$  and parametrise  
   \[   {\cal C} = \hat{g}^{-1} \bar{\nabla} \hat{g}  + \hat{g}^{-1}  {\cal C}^\prime  \hat{g} \, . \]  Acting on the field ${\cal C'}, {\cal B} $ and $\hat{g}$ we inherit both the original gauge symmetries and new internal symmetries refelcting the rundancy of the parameterisation.  These act as:
\begin{align}\nonumber
    \hat{f}: &\quad  &{\cal B}  \mapsto {\cal B}  \, , \quad \, & {\cal C}^\prime   \mapsto {\cal C}^\prime,   \, \quad &\hat{g} \mapsto \hat{g}\hat{f} \, ,\\ \nonumber 
    \check{f}:& \quad & {\cal B}  \mapsto {\cal B}  \, , \quad \,  &{\cal C}^\prime   \mapsto \check{f}^{-1} \bar{\nabla} \check{f}  + \check{f}^{-1}  {\cal C}^\prime \check{f} ,  \, \quad &\hat{g} \mapsto \check{f}^{-1}\hat{g} \, , \\ \nonumber
    \hat{h}: &\quad   &{\cal B}  \mapsto \hat{h}^{-1}\bar{\partial} \hat{h}+ \hat{h}^{-1} {\cal B} \hat{h}  \, , \quad \,  &{\cal C}^\prime   \mapsto \hat{h}^{-1}  {\cal C}^\prime \hat{h}    , \, \quad & \hat{g} \mapsto \hat{h}^{-1} \hat{g} \hat{h} \, . \nonumber
\end{align}
\item {\em Set boundary conditions} We impose, by choice, a set of boundary conditions on ${\cal C}$ that ensure the vanishing of   
\[ 
\frac{1}{2\pi i}\int _{\bPT} \bar{\partial} \Omega\wedge\tr\,{\cal C} \wedge   \delta {\cal C}\, .
\]
Structurally this has the same form as in the ungauged theory, so whatever boundary conditions were employed on ${\cal A}$ in that case could equally be applied for ${\cal C}$ in this gauged version. 
Our gauge redundancies are reduced to those that preserve the chosen set of boundary conditions
\item {\em Partial Fix} We use the $\hat{h}$ symmetry and the internal $\check{f}$ symmetry to make the fixing that ${\cal C}'_0=0$ and ${\cal B}_0= 0$,  i.e. that neither field has $\mathbb{CP}^1$ legs. As in the ungauged case, this gauge choice is consistent with the constraints.  In bulk, away from the poles, this fixing requires the equations for the remaining components are pure constraint and that 
\[{\cal C}^\prime_{\dot{a}} = \pi^a C_{a \dot{a} } \, , \quad {\cal B}_{\dot{a}} = \pi^a B_{a \dot{a} }\,.
   \]
Note that the bulk terms then vanish upon these constraints so that the theory reduces simply to 
\begin{equation}\label{6d_action_in_prime}
\frac{1}{2\pi i}\int _{\bPT} \bar{\partial} \Omega\wedge      \mathrm{tr}  \left({\cal C}^\prime  \wedge  \bar{\nabla} \hat{g} \hat{g}^{-1} - \hat{g}^{-1}  \bar{\partial}  \hat{g} \wedge \cB- \bar{\partial} \hat{g} \hat{g}^{-1}\wedge \cB -  \cB  \wedge \hat{g} \cB  \hat{g}^{-1}  \right)  -\Omega \wedge {\cal L}_{\text{WZ}}[\hat{g}]\,.
\end{equation}

In addition we may further choose to fix some properties on the edge mode $\hat{g}$, the precise details of which would be dependent on the choice of theory, but again we emphasise that the analysis in the ungauged model would be equally valid in this.   

\item {\em Determine $C_{a \dot{a}}$}   We now solve the boundary conditions to determine $C_{a \dot{a}}$ in terms of $\hat{g}$ (or rather its evaluation at the poles) and $B_{a \dot{a}}$.     Indeed, if this analysis has been performed already in the ungauged model, it is now manifest that the result will remain the same up to the modification that partial derivatives become covariantised with respect to the, now 4d, gauge field $B_{a \dot{a}}$.

\item {\em Localise} We are now in a position to complete the localisation, by inserting into the edge mode action the solution for ${\cal C'}$ in term of $\hat{g}$ and $B$, evaluating ${\bar \partial} \Omega$ to produce delta functions and their derivatives at poles allowing the   integration over the $\mathbb{CP}^1 $  directions to take place.   Thus, modulo the straightforward covariantisation by 
$\cB$, the gauged construction inherits the localisation structure of the ungauged model, paving the way to concrete examples such as the 
(1,1,2) pole structure relevant to the $\lambda$-deformation of gauged WZW models that we now consider.
\end{enumerate}

\section{Example $\lambda$-model }

In this section we will utilize the above formulation of holomorphic Chern-Simons theory on $\bPT$ with the meromorphic $(3,0)$-form relevant for the $\lambda$-model, as per equation \eqref{112_pole_structure}.   As this is a technical section let us direct the reader to the main results: the presentation of the 4d integrable model, eq.~\eqref{4d_action}; the establishment of its equations of motion as \\(anti-)self-duality of a connection $A$;  the reduction to a gauged two-field 2d integrable model, eq.~\eqref{eq:Action2d} and its Lax formulation, eq.~\eqref{eq:lax_components_from_6d}; and the identification of a parametric point of enhanced symmetry resulting in the simpler 2d integrable model of eq.~\eqref{eq:Action2special}.

According to our general discussion we are going to employ the boundary conditions used in the ungauged model  \cite{Cole:2023umd}.  These conditions involve cancellations between contributions at the poles $\pi =\alpha$ and $\pi = \tilde{\alpha}$ and are expressed in terms of a free scale parameter $\sigma$, and a unit norm spinor $\mu$:  
\begin{equation}\label{eq:bcugly}
\begin{split}
  &  \left. \frac{[\cC\hat\mu]}{\langle\alpha\beta\rangle}\right|_{\alpha}=\sigma^{-1}\left.\frac{[\cC\hat\mu]}{\langle\tilde\alpha\beta\rangle}\right|_{\tilde{\alpha}},\quad \left.\frac{[\cC\mu]}{\langle\alpha\beta\rangle}\right|_{\alpha}=\sigma\left.\frac{[\cC\mu]}{\langle\tilde\alpha\beta\rangle}\right|_{\tilde{\alpha}},\quad\cC\big|_{\beta}=0 \,.
    \end{split}
\end{equation}
To elucidate these conditions we work with  adapted  coordinates   
\newcommand{\wsf}{\mathsf{w}}
\newcommand{\zsf}{\mathsf{z}}
\newcommand{\hwsf}{\hat{ \mathsf{w} }}
\newcommand{\hzsf}{\hat{ \mathsf{z} }}
\begin{equation}\begin{aligned}\label{eq:newcoordinates}
\wsf & = \frac{\langle\alpha\beta\rangle}{\langle\alpha\tilde\alpha\rangle}\tilde \alpha_{a} x^{a \dot{a}}\hat\mu_{\dot a }   ~,
\qquad &
\hwsf & = -\frac{\langle\alpha\beta\rangle}{\langle\alpha\tilde\alpha\rangle} \tilde \alpha_{a} \mu_{\dot a}x^{a\dot a} ~,
\\
\zsf & = -\frac{\langle\tilde\alpha\beta\rangle}{\langle\alpha\tilde\alpha\rangle}  \alpha_{a}\hat\mu_{\dot a} x^{a \dot a} ~, \qquad &
\hzsf & = \frac{\langle\tilde\alpha\beta\rangle}{\langle\alpha\tilde\alpha\rangle} \alpha_{a}\mu_{\dot a} x^{a\dot a} ~,  
\end{aligned}\end{equation}
such that 
     \begin{equation}\label{volume_form_in_sf}
\mathrm{d}\wsf\wedge\mathrm{d}\hwsf\wedge\mathrm{d}\zsf\wedge\mathrm{d}\hzsf= \frac{\langle\alpha\beta\rangle^2\langle\tilde\alpha\beta\rangle^2}{\langle\alpha\tilde\alpha\rangle^2}\mathrm{vol}_4 \, .\nonumber
 \end{equation}
Recall that as a hyperk\"{a}hler manifold, $\mathbb{R}^4$ is equipped with a $\mathbb{CP}^1$'s worth of complex structures. The coordinates $\zsf $ and $\hzsf$ are holomorphic with respect to the complex structure defined by the point $\pi = \alpha$ and $\wsf$ and $\hwsf$ by the one at $\pi = \tilde{\alpha}$.   Given a complex structure defined by $\pi\in \mathbb{CP}^1$,    the self-dual two-form of type $(2,0)$   is defined by $\Sigma_\pi = \pi_a \pi_b \epsilon_{\dot{a} \dot{b} } \dr x^{a \dot{a}} \wedge \dr x^{b \dot{b}}$ and in particular we have 
\newcommand{\usf}{\mathsf{u}}
\newcommand{\vsf}{\mathsf{v}}
\newcommand{\husf}{\hat{ \mathsf{u} }}
\newcommand{\hvsf}{\hat{ \mathsf{v} }}
\begin{equation}
    \Sigma_{\beta } = 2 \dr \usf \wedge \dr \husf  \, ,\quad \Sigma_{\alpha } = 2 \frac{\langle \alpha \tilde{\alpha}\rangle^2}{\langle   \tilde{\alpha} \beta \rangle^2 }\dr \zsf \wedge \dr \hzsf \, ,\quad \Sigma_{\tilde{\alpha}} = 2 \frac{\langle \alpha \tilde{\alpha}\rangle^2}{\langle    \alpha \beta \rangle^2 }\dr \wsf \wedge \dr \hwsf  \, ,  \nonumber
\end{equation} 
 in which  $\usf=  \zsf +   \wsf$, $\husf=  \hzsf +   \hwsf$ (we will also make use occasionally of $\vsf = \wsf - \zsf$ and $\hvsf = \hwsf - \hzsf$). 

Expressing ${\cal C} = {\cal C}_{\dot{a}} \bar{e}^{\dot{a}}$ in terms of these coordinates and using the localisation methods that we shall detail momentarily, we find that the boundary term in the variation of the action includes contributions at the single poles of the form 
 $$\int_{\mathbb{R}^4} \textrm{vol}_4 \,  \left(  \iota_{  \wsf} {\cal C} \iota_{\hwsf }  \delta  {\cal C} -\iota_{  \hwsf} {\cal C} \iota_{\wsf }  \delta  {\cal C}     \right) \vert_\alpha -  \left(  \iota_{  \zsf} {\cal C} \iota_{\hzsf  }  \delta  {\cal C}  -\iota_{  \hzsf} {\cal C} \iota_{\zsf   }  \delta  {\cal C}     \right) \vert_{\tilde{\alpha}} \, , 
 $$ 
 in which we denote, for example, $\iota_{\wsf} \equiv \iota_{\partial_{\wsf}} $ the contraction with respect to $\partial_{\wsf}$.   Thus we see that  the boundary conditions at the single poles, eq.~\eqref{eq:bcugly},  which read in these coordinates 
\begin{equation}\label{BC_4d_fields}
  \iota_{\wsf} {\cal C} \vert_{\alpha }  = \sigma \iota_{\zsf}  {\cal C} \vert_{\tilde{\alpha} } \ , \quad   \iota_{\hwsf}  {\cal C}\vert_{\alpha }  = \sigma^{-1 }   \iota_{\hzsf}  {\cal C} \vert_{\tilde{\alpha} }\, ,\nonumber
\end{equation}
do indeed set to zero the contribution above.  Of particular note is that freedom to include the constant parameter $\sigma$ such that this defines an entire family of possible models.  

  We now express these boundary conditions in terms of the Stueckelberg parameterisation 
\begin{equation} {\cal C} = \hat{g}^{-1} \bar{\nabla} \hat{g}  + \hat{g}^{-1}  {\cal C}^\prime  \hat{g}  \, ,   \nonumber
\end{equation}
under the gauge fixing that ${\cal C}^\prime_0 = 0  $  subject to the fixing and definitions 
\begin{equation}
    \hat{g} \vert_\alpha = g \, , \quad \hat{g} \vert_{\tilde{\alpha}}= \tilde{g}   \, , \quad \hat{g} \vert_\beta=  \text{id}_G  \, . \nonumber
\end{equation}
 Given the bulk requirement that  ${\cal C}^\prime_{\dot{a}} = \pi^a C_{a \dot{a}}$ we can address the boundary condition at the double pole in terms of the 4d gauge field $C = C_{a \dot{a}} \dr x^{a \dot{a}}$:
\begin{equation}
 {\cal C}\vert_{\beta}= 0 \Leftrightarrow   {\cal C}^\prime \vert_{\beta}= 0 \Leftrightarrow  \beta^{a} C_{a \dot{a}} = 0 \Leftrightarrow  C_{a \dot{a}} \dr x^{a \dot{a}}  = C_{\usf} \dr \usf   +  C_{\husf} \dr \husf \, . \nonumber
\end{equation}
The boundary conditions at the first order poles then become 
\begin{equation}\label{eq:bcredux}
g^{-1}\nabla_{\wsf}  g + g^{-1} C_{\usf} g = \sigma \left(\tilde{g}^{-1}\nabla_{\zsf} \tilde{g} +\tilde{g}^{-1} C_{\usf}\tilde{g} \right), \quad g^{-1}\nabla_{\hwsf}  g + g^{-1} C_{\husf} g =  \sigma^{-1}\left(\tilde{g}^{-1}\nabla_{\hzsf}\tilde{g}   +\tilde{g}^{-1} C_{\husf}\tilde{g} \right) \, .
\end{equation}

Using the short hand notation $R^\nabla  =\nabla g g^{-1}$ and $\tilde{R}^\nabla  =\nabla\tilde{g}\tilde{g}^{-1}$ 
we can hence determine
\begin{equation}
    \label{eq:CinG}
C_{\usf}  = - U_+ R^\nabla_\wsf - U_-^T \tilde{R}^\nabla_\zsf \, , \quad C_{\husf} =  - U_- R^\nabla_{\hwsf} - U_+^T \tilde{R}^\nabla_{\hzsf} \, ,
\end{equation} 
where the operators $U_{\pm}$ are defined  as 
$$
  U_\pm = (\textrm{id} - \sigma^{\pm 1} \Ad_{g} \circ \Ad_{\tilde{g}}^{-1}  )^{-1} \,  .
 $$

\subsection{Localisation}
We now turn to the final step of the procedure described above, namely performing the integral over the $\mathbb{CP}^1$.   In general we are faced with integrals of the form  
\begin{equation}\label{General form of integral}
    I = \frac{1}{2 \pi i} \int_{\bPT} \bar{\pd} \Omega \wedge Q \, , 
\end{equation} 
where in general  $  Q \in \Omega^{(0,2)} (\bPT)$ however since the legs in the $\bCP^{1}$ directions are saturated by the integration measure $\bar{\pd} \Omega$ we can restrict to consider $Q = Q_{\dot{a} \dot{b}} \bar{e}^{\dot{a}} \wedge \bar{e}^{\dot{b}}$.
To understand these it is  expedient to move to inhomogeneous coordinates on $\mathbb{CP}^1$ setting $\pi = (1, \zeta) $  
such that we can employ the identities
\begin{equation} \label{eq:inhom_pole_der}
    \pd_{\bar{\zeta}} \bigg( \frac{1}{\zeta - \alpha} \bigg) = - 2 \pi \rmi \, \delta^{2}(\zeta - \alpha) \ , \qquad 
    \int_{\bCP^{1}} \dr \zeta \wedge \dr \bar{\zeta} \, \delta^{2}(\zeta - \alpha) \, f(\zeta) = f(\alpha) \ . \nonumber
\end{equation}
Having done this one can localise $I$ to a four-dimensional integral, and revert back to homogeneous spinor coordinates.    
For the case at hand of $\Omega$  given in eq.~\eqref{112_pole_structure}, we have the general result  

\begin{align}\label{formula_1}
        I =  \frac{K}{4}\int_{\mathbb{R}^4}  & 
\frac{1}{\langle \alpha \tilde{\alpha} \rangle \langle \alpha \beta  \rangle^2     }  \Sigma_{\alpha }\wedge Q\vert_{ \alpha   }  -  \frac{1}{\langle \alpha \tilde{\alpha} \rangle \langle \tilde{\alpha } \beta  \rangle^2     }  \Sigma_{\tilde{\alpha }   }\wedge Q\vert_{\tilde{\alpha } } \nonumber \\ 
& +\frac{ \left( \langle\alpha\beta\rangle\langle\tilde{\alpha} \hat{\beta } \rangle  + \langle\alpha\hat{\beta} \rangle\langle\tilde{\alpha}  \beta \rangle\right)   }{\langle \alpha\beta \rangle^2  \langle \tilde{\alpha}\beta \rangle^2 \langle \beta \hat{\beta} \rangle   } \Sigma_{\beta  } \wedge Q\vert_\beta    - \frac{1}{\langle \alpha \beta\rangle \langle \tilde{\alpha} \beta\rangle 
 }  \Sigma_{\beta } \wedge \pd_0Q \vert_\beta \, .\nonumber 
\end{align} 
When $Q = \pi^{a} \pi^{b} q_{a \dot{a} b \dot{b} } \bar{e}^{\dot{a}} \wedge \bar{e}^{\dot{b}}  $    we have the further simplifications 
$$
\Sigma_{\pi }\wedge Q = \Sigma_{\pi }\wedge q \, ,  \quad \Sigma_{\pi  }\wedge \partial_0 Q = \Sigma_{\pi }\wedge \pd_0q + \frac{2}{\langle\pi \hat{\pi} \rangle } \xi_\pi \wedge q    \, ,   
$$  
in which $ q= q_{a \dot{a} b \dot{b} }  \dr x^{a\dot{a} }\wedge   \dr x^{b\dot{b}}$ and $\xi_\pi  =  \pi_a \hat{\pi}_b \epsilon_{\dot{a} \dot{b} } \dr x^{a \dot{a}} \wedge \dr x^{b \dot{b}}  $ .

We begin with the contribution to \eqref{6d_action_in_prime} in which 
\begin{equation} 
Q_1   =  \tr\,  {\cal C}^\prime \wedge\bar{\nabla} \hat{g} \hat{g}^{-1} \, \quad \textrm{such that}   \quad q_1 =\tr \,C \wedge (\dr \hat{g} \hat{g}^{-1} + B - \hat{g} B \hat{g}^{-1})    \, . 
 \nonumber
\end{equation}
Since  the boundary condition ensures ${\cal C}'  \vert_\beta  = \partial_0 {\cal C}' \vert_\beta  = 0 $ we find, after some moderate algebra using the solution for the components of $C_{a \dot{a}}$ eq.~\eqref{eq:CinG}, the localised contribution yields
\begin{equation}
    \begin{aligned}
     I_1  &= \frac{1}{2 \pi i} \int_{\bPT} \bar{\pd} \Omega \wedge Q_1  \\ 
 &=-\frac{K}{\langle\alpha\tilde{\alpha}\rangle}\int _{\mathbb{R}^4}\mathrm{vol}_4\, \tr \Big[\frac{1}{2}R^\nabla_\wsf (U_{+}^T-U_{-})R^\nabla_{\hwsf}+\frac{1}{2}\tilde{R}^\nabla_{\zsf}(U_{+}^T-U_{-})\tilde{R}^\nabla_{\hzsf}-R^\nabla_\wsf U_{+}^T\tilde{R}^\nabla_{\hzsf}+\tilde{R}^\nabla_{\zsf}U_{-}R^\nabla_{\hwsf} \Big]\, .\nonumber 
\end{aligned} 
\end{equation}
This contribution matches the ungauged model with the replacement of derivatives to  derivatives covariantised with the connection $B$. 
The next contribution comes from the WZ terms, and here we explicitly introduce an extension parametrised by coordinate $\rho$, 
$$
Q_2 =- \int _I\dr \rho \;  \tr \left( \hat{g}^{-1}\partial_\rho \hat{g}\,   \hat{g}^{-1}\bar{\pd } \hat{g} \wedge  \hat{g}^{-1}  \bar{\pd }  \hat{g} \right) \, ,
$$
which gives
 \begin{equation}
    I_2  = -\frac{K}{ 2\langle \alpha \tilde{\alpha} 
\rangle}  \int_{\mathbb{R}^4\times I } \textrm{vol}_4\wedge \dr \rho \,    \tr \left( g^{-1} \partial_\rho g [ g^{-1} \partial_\wsf g  ,  g^{-1} \partial_{\hwsf} g   ]  - \tilde{g}^{-1} \partial_\rho \tilde{g}[ \tilde{g}^{-1} \partial_\zsf \tilde{g}  ,  \tilde{g}^{-1} \partial_{\hzsf}\tilde{g}  ] \right)   \, .\nonumber
 \end{equation}
The final term to consider is the gauge completion of the WZ contribution given by  
$$
Q_3 = -\tr \left( \hat{g}^{-1}  \bar{\partial}  \hat{g} \wedge \cB+\bar{\partial} \hat{g} \hat{g}^{-1}\wedge \cB + \cB  \wedge \hat{g} \cB  \hat{g}^{-1}  \right) \,. 
$$
This term is more delicate as one needs to carefully consider contributions from all poles, including the double pole at $\beta$.   At the single poles we have   
$$
q_3 \vert_\alpha = - \tr \left(   g^{-1} \dr g\wedge B +  \dr g g^{-1}\wedge  B +  B  \wedge g B  g^{-1}  \right) \, , 
$$
and similarly a contribution for $\tilde{g}$ arises at $\tilde{\alpha}$.
These first order pole contributions  combine with the WZ terms produced in $I_2$ above to yield the gauged WZ Lagrangian.   

At the double we need to consider the Lie algebra valued field,  $u \propto  \hat{g}^{-1} \partial_0 
\hat{g} \vert_\beta $ (where we choose a normalisation of $u$ of convenience to absorb factors) 
$$
q_3 \vert_\beta =0 \, , \quad \partial_0 q_3\vert_\beta  = \tr( u F[B]   + \dr ( u B)) \, .
$$
Hence the combined contribution from $I_2$ and $I_3$ can be expressed as  (omitting the total derivative $\mathrm{d}(uB)$ assuming appropriate fall-off of the fields on $\mathbb{R}^4$) 
\begin{equation}
\begin{split}
    I_2 + I_3 =&   -\frac{K}{ 2\langle\alpha\tilde\alpha\rangle}    \int \mathrm{vol}_4( \iota_{\hwsf} \iota_{\wsf} \mathcal{L}_{\text{gWZ }}[g,B]  - \iota_{\hzsf} \iota_{\zsf} \mathcal{L}_{\text{gWZ }}[\tilde{g},B] )  )\\
    & - \frac{K}{ \langle\alpha\tilde\alpha\rangle} \int_{\mathbb{R}^4} \mathrm{d}\usf \wedge \mathrm{d}\husf\wedge \Tr(u F[B])   \, ,
    \end{split}\nonumber
\end{equation}
 where we recall the gauged WZW Lagrangian two-form is
\begin{equation}\label{LgWZ_2form} \mathcal{L}_{\text{gWZ}}[g,B]=\tr\left(g^{-1}\mathrm{d}g\wedge B+\mathrm{d}g g^{-1}\wedge B + B\wedge g B g^{-1}\right) + \frac{1}{3}\int_{I}\mathrm{d}\rho\,\tr \,g^{-1} \partial_\rho g\, g^{-1}\mathrm{d}g\wedge g^{-1}\mathrm{d}g \,.
\end{equation}

Combining all the terms produces we recover the action, 
  \begin{equation}\label{4d_action}
      \begin{split}
          S_{\text{IFT}_4}&=-\frac{K}{\langle\alpha\tilde{\alpha}\rangle}\int_{\mathbb{R}^4}\mathrm{vol}_4\,\tr \Big\{ \frac{1}{2}R^{\nabla}_\wsf(U_{+}^T-U_{-})R^{\nabla}_{\hwsf}+ \frac{1}{2}\tilde{R}^{\nabla}_\zsf(U_{+}^T-U_{-})\tilde{R}^{\nabla}_{\hzsf}-R^{\nabla}_{\wsf}U^{T}_{+}\tilde{R}^{\nabla}_{\hzsf}+\tilde{R}^{\nabla}_{\zsf}U_{-}R^{\nabla}_{\hwsf}\\
          &\hspace{3.6cm}+\frac{1}{2}\left( \iota_{\hwsf} \iota_{\wsf} \mathcal{L}_{\text{gWZ }}[g,B]  - \iota_{\hzsf} \iota_{\zsf} \mathcal{L}_{\text{gWZ }}[\tilde{g},B]\right)\Big\}\\
            & \hspace{0.6cm}-\frac{K}{ \langle\alpha\tilde\alpha\rangle} \int_{\mathbb{R}^4} \mathrm{d}\usf \wedge \mathrm{d}\husf\wedge \tr(u F[B]) \, .
      \end{split}
  \end{equation}
 \subsection{Equations of motion}
As detailed in appendix \ref{appendix_EOM}, the equations of motion of the action eq.~\eqref{4d_action}  are given by 
\begin{align*}
    \delta g: &\quad 0  = F^\nabla_{\wsf\hwsf}[C] + F_{\wsf\hwsf}[B] \, ,\\
        \delta \tilde{g}: &\quad 0  = F^\nabla_{\zsf\hzsf}[C] + F_{\zsf\hzsf}[B] \, ,\\
        \delta u: &\quad 0 = \dr \usf \wedge \dr \husf \wedge F[B]\, ,
\end{align*}
in which we use  
$$
F^\nabla[C] = \nabla C +  C \wedge C \, ,\quad F[B] = \dr B + B \wedge B\, ,
$$
with $C$ is evaluated in terms of the currents in eq.~\eqref{eq:CinG}.    Recalling that $C = C_{\husf } \dr \husf + C_\usf  \dr \usf   $ we also have identically $ \dr \usf \wedge \dr \husf  \wedge F^\nabla[C]  = 0 $, and thus these equations combine to the anti-self-duality of $F^\nabla[C] + F[B]$.    Indeed, if we let $C= A - B$ (now working in terms of the four-dimensional connections) these equations correspond exactly to the anti-self-duality of $A$.

Turning to the equations of motion that follow from variation with respect to $B$ we obtain
\begin{align}
    \delta B_{\wsf}: &\quad  0  =   Q_{\hwsf} - \nabla_{\hvsf} u  \, , \quad 
        \delta B_{\hwsf}: \quad  0  =   Q_\wsf - \nabla_\vsf u  \, , \nonumber \\ 
        \delta B_{\zsf}: &\quad  0  = \widetilde{Q}_{\hzsf} - \nabla_{\hvsf} u \, ,\quad 
        \delta B_{\hzsf}: \quad  0  = \widetilde{Q}_{\zsf} - \nabla_{\vsf} u   \, . \nonumber
\end{align}
in which we define  
$$
Q = C- C^g  \vert_\fh   =  A - A^g \vert_\fh \, ,\quad \widetilde{Q} =  C- C^{\tilde{g}}  \vert_\fh  =    A - A^{\tilde{g}}  \vert_\fh \, ,
$$ 
with $A^g = g^{-1} \dr g + g^{-1} A g  $ and $C^g = g^{-1} \nabla g + g^{-1} C g $.  
Combining these yields a condition 
\begin{equation}
Q_{\wsf} - \widetilde{Q}_{\zsf} = 0 \, , \quad Q_{\hwsf} - \widetilde{Q}_{\hzsf} = 0 \, .\nonumber
\end{equation}
Note that the construction of $C$ that arose in eq.~\eqref{eq:bcredux} ensures that 
\begin{equation}\label{eq:bcreduxv2}
C_\wsf = C_\zsf \, ,\quad   C^g_\wsf = \sigma  C^{\tilde{g}}_\zsf \,  ,  \quad 
C_{\hwsf} = C_{\hzsf} \, , \quad  C^g_{\hwsf} = \sigma^{-1} C^{\tilde{g}}_{\hzsf} \,   . 
\end{equation}
and hence we obtain the condition   
\begin{equation}
(1- \sigma ) C_\zsf^{\tilde{g}} \vert_\mathfrak{h}  = 0 \, , \quad (1- \sigma^{-1} ) C_{\hzsf}^{\tilde{g}} \vert_\mathfrak{h}  = 0 \, ,\nonumber
\end{equation}
 or explicitly 
$$ (1 - \sigma^{-1}) \Ad_{\tilde{g}}^{-1} U_+( \tilde{R}^\nabla_{\zsf} - R^\nabla_{\wsf})  \vert_\fh = 0 \, , \quad (1 - \sigma ) \Ad_{\tilde{g}}^{-1} U_-( \tilde{R}^\nabla_{\hzsf} - R^\nabla_{\hwsf})  \vert_\fh  = 0 \, . $$ 
Since our construction assumes $\sigma \neq 1 $, (else the matrices $U_\pm$ are ill defined) these conditions invoke the vanishing of the subgroup components of the currents, and this determine implicitly  two of four components of $B$ in terms of the other fields in the model.   

Given that $F_{\vsf \hvsf}[B] = 0 $ by virtue of the $u$ equation of motion, we obtain an integrability condition that 
\begin{equation}
\dagger \equiv \nabla_\vsf  Q_{\hwsf} - \nabla_{\hvsf} Q_{\wsf} = F_{\vsf \hvsf}[B]\cdot u =   0 \, ,\nonumber
\end{equation} 
which might, in principle, be a cause for concern as it could impose conditions on $g$ and $\tilde{g}$ on top of the anti-self-duality of $A$.  However with some work we can show that this integrability condition is automatically satisfied on the remaining equations of motion.    First we use the $B$ equation of motion   $Q_\wsf = \widetilde{Q}_\zsf$  to recast as 
$$
\dagger = \iota_\wsf \iota_{\hwsf} \nabla (C -  C^g )\vert_\frak{h} - \iota_\zsf \iota_{\hzsf}\nabla ( C -   C^{\tilde{g}} )\vert_\frak{h}\, . 
$$
Now if $\frak{g}= \frak{h} \oplus \frak{m}$ is a reductive decomposition, which we will assume, then $[\frak{h} , \frak{m}] \subset \frak{m}$ and   we can take the projection after the covariant derivative i.e.  
$$
\dagger = \iota_\wsf \iota_{\hwsf} (\nabla C - \nabla  C^g )\vert_\frak{h} - \iota_\zsf \iota_{\hzsf}(\nabla  C -  \nabla  C^{\tilde{g}} )\vert_\frak{h}\, . 
$$ 
To complete these derivatives of $C$ into the combination $F^\nabla[C]$ we are required to incorporate contributions of the type $C \wedge C$.  Fortunately the conditions in eq.~\eqref{eq:bcreduxv2} ensure 
$$ 
 \iota_\wsf \iota_{\hwsf} C\wedge C =  \iota_\zsf \iota_{\hzsf} C\wedge C  \, , \quad  \iota_\wsf \iota_{\hwsf} C^g\wedge C^g =  \iota_\zsf \iota_{\hzsf} C^{\tilde{g}}\wedge C^{\tilde{g}}  \, . 
$$
and thus we can recast the integrability condition as   
$$
\dagger= \iota_\wsf \iota_{\hwsf} (F^\nabla[C] -F^\nabla[ C^g]   )\vert_\frak{h} - \iota_\zsf \iota_{\hzsf}(F^\nabla[C] -F^\nabla[ C^{\tilde{g}}]   ) \vert_\frak{h}\,.
$$
Now we can use  $F^\nabla[C^g] = \Ad_g^{-1} F^\nabla[C] + \Ad_g^{-1} F[B] - F[B]$, which follows directly from the definition, to obtain 
$$
\dagger = (1- \Ad_g^{-1}) \iota_\wsf \iota_{\hwsf}   (F^\nabla[C] + F[B])\vert_\frak{h} - ( 1- \Ad_{\tilde{g}}^{-1}) \iota_\zsf\iota_{\hzsf}( F^\nabla[C] + F[B]  ) \vert_\frak{h} \, ,
$$ 
which immediately now vanishes upon  the $g,\tilde{g}$ equation of motion.    

\subsection{Reduction}\label{section:reduction}

We now perform a reduction to two dimensions.  The data that defines the reduction consists of a set of unit norm spinors $\gamma_a$ and $\kappa_{\dot{a}}$ for which we use holomorphic coordinates 
\newcommand{\zz}{\mathtt{z}}
\newcommand{\zzb}{\bar{\mathtt{z}} }
\newcommand{\ww}{\mathtt{w}}
\newcommand{\wwb}{\bar{\mathtt{w}} }
\begin{equation}
    \dr \zz = \langle \gamma | \dr x | \kappa]  \, ,\quad  \dr \zzb = \langle \hat\gamma | \dr x | \hat\kappa]  \, ,\quad  \dr \ww = \langle \gamma | \dr x | \hat \kappa]  \, ,\quad  \dr \wwb = - \langle \hat\gamma | \dr x |   \kappa] \, .\nonumber
\end{equation} 
To convert the coordinate differentials defined in eq.~\eqref{eq:newcoordinates}  we express spinors in this basis as 
$$
\pi = \langle \pi \hat\gamma  \rangle \gamma - \langle \pi  \gamma  \rangle \hat \gamma \, , \quad \mu =\left[   \mu\hat \kappa   \right] \kappa   - \left[   \mu \kappa    \right] \hat \kappa \, . 
$$
We will in this work make the choice $\mu = \kappa$, i.e. to align the reduction directions with the choice of dotted spinor used to define the boundary conditions of the six-dimensional parent theory.\footnote{This choice leads to many simplifications and in particular leads to a 2d invariant theory; other choices of $\kappa$ can be considered and lead to quite intricate actions that do not hold sigma-model interpretations.  However the form of the Lax for the generic choice of $\kappa$ remains relatively elegant and is presented in Appendix \ref{general_reductions}.} The details about the coordinate transformation and full form of the action \eqref{4d_action} in terms of the $(\zz,\zzb,\ww,\wwb)$ coordinates are contained in Appendix \ref{appendix_4d_action}.

We proceed by using a reduction ansatz of the form 
\begin{equation}
\label{eq:RedAnsSimple}
\pd_\zz = \pd_{\zzb} = 0 \, \quad B_\zz=  0 \, , \quad B_{\zzb} = 0 \, , 
\end{equation}
such that anything with a $\zz$ or $\zzb$ index in \eqref{4d_action_in_new_coords} is set to zero. The result of this depends naturally on combinations

\begin{equation}
k = K  \frac{  \langle\alpha \gamma \rangle  \langle\alpha\hat\gamma\rangle  }{\langle\alpha\tilde\alpha\rangle  \langle\alpha\beta \rangle^2    }  \, , \quad \tilde{k} = K  \frac{  \langle \tilde \alpha \gamma \rangle  \langle \tilde \alpha\hat\gamma\rangle  }{\langle\alpha\tilde\alpha\rangle  \langle\tilde  \alpha\beta \rangle^2    }   \, , \quad \mathtt{t}^2 =\frac{\langle\alpha \hat\gamma \rangle \langle \tilde\alpha  \gamma \rangle }{ \langle\alpha  \gamma \rangle   \langle \tilde\alpha  \hat \gamma   \rangle } \, ,\nonumber
\end{equation}
and is given by
\begin{equation}\label{eq:Action2d}
    \begin{split}
        S_{\text{IFT}_2}=-\int_{\mathbb{R}^2} \mathrm{d}\ww\wedge\mathrm{d}\wwb&\,\tr\, \Big\{\frac{k}{2}  R^{\nabla}_{\ww}(U_{+}^{T}-U_{-})R^{\nabla}_{\wwb} + \frac{\tilde{k}}{2} \widetilde{R}^{\nabla}_{\ww}(U_{+}^{T}-U_{-}) \widetilde{R}^{\nabla}_{\wwb}   \\
        &+ \sqrt{k \tilde{k}} \Big[  - \mathtt{t}^{-1}  R^\nabla_\ww U_+^T \widetilde{R}^\nabla_{\wwb}  +  \mathtt{t}\,    \widetilde{R}^\nabla_\ww U_- R^\nabla_{\wwb}    \Big]\Big\} \\ 
     &  + \int_{\mathbb{R}^2} \tr (u'  F[B]) - \frac{k}{2}  {\cal L}_{\text{gWZ}}[g,B]+ \frac{\tilde{k }}{2}  {\cal L}_{\text{gWZ}}[\tilde{g},B] \, ,
        \end{split}
\end{equation}
where for aesthetic reason we have rescaled the Lagrange multiplier 
$ 
\langle\alpha\tilde\alpha\rangle u' =  K \langle\beta\gamma\rangle \langle\beta\hat\gamma\rangle u   
$. 
The  equations of motion (see appendix~\ref{sec:2deqm} for details)  of the above reduced action are 
\begin{equation}\label{EOM_g_2d}
\delta g : \quad    \nabla_{\ww} V_{\wwb} -  \nabla_{\wwb} V_{\ww} - \frac{1}{k} [V_{\ww} ,V_{\wwb}  ]   =    k F[B]_{\ww\wwb} \, ,
\end{equation}
\begin{equation} 
\delta \tilde{g}:    \nabla_{\ww}  \widetilde{ V}_{\wwb} -  \nabla_{\wwb}  \widetilde{ V}_{\ww} - \frac{1}{\tilde{k}}  [\widetilde{ V}_{\ww} ,\widetilde{ V}_{\wwb}  ]   =      \tilde{k}  F[B]_{\ww\wwb} \, ,\nonumber
\end{equation}
\begin{equation} 
\delta u':     0 =   F[B]_{\ww\wwb} \, ,\nonumber
\end{equation}
in which we have introduced currents  
\begin{equation}\label{define_currents}
    V_{\ww } = k U_+ R^\nabla_{\ww}  + \sqrt{\tilde{k} k}  \texttt{t}    U_-^T\widetilde{R}^\nabla_{\ww} \, , \quad V_{\wwb } = k U_- R^\nabla_{\wwb}  + \sqrt{\tilde{k} k}  \texttt{t}^{-1}    U_+^T\widetilde{R}^\nabla_{{\wwb}}  \, ,
\end{equation}
\begin{equation}
\widetilde{ V}_{\ww} = \sqrt{\frac{\tilde{k}}{k}} \texttt{t}^{-1} V_{\ww }
 \, , \quad \widetilde{ V}_{\wwb} = \sqrt{\frac{\tilde{k}}{k}} \texttt{t} V_{\wwb }  \,  \nonumber . 
 \end{equation}
To make direct contact with the Lax connection obtained from the six-dimensional twistorial perspective it is useful to introduce inhomogenous coordinate representatives of the points in $\mathbb{CP}^1$: 
\renewcommand{\aa}{a}
\renewcommand{\bb}{b}
\newcommand{\pp}{\zeta}
\newcommand{\aat}{\tilde{a} } 
\begin{equation}
\aa  = \frac{ \langle \alpha \gamma \rangle  }{ \langle \alpha  \hat{\gamma } \rangle  } \, , \quad \aat  = \frac{ \langle \tilde\alpha \gamma \rangle  }{ \langle \tilde\alpha  \hat{\gamma } \rangle  } \, , \quad \bb   = \frac{ \langle \beta  \gamma \rangle  }{ \langle  \beta   \hat{\gamma } \rangle  } \, , \quad \pp    = \frac{ \langle \pi \gamma \rangle  }{ \langle  \pi   \hat{\gamma } \rangle  }  \, .\nonumber
\end{equation}
Note that 
\begin{equation}
\frac{k}{\tilde{k}} = \frac{\aa (\aat - \bb)^2  }{\aat (\aa - \bb)^2 } \, , \quad \mathtt{t}^2 = \frac{\aat}{ \aa }\, ,\nonumber
\end{equation}
and that what enters in the equation of motion are exactly these ratios rather than $k$ and $\tilde{k}$ separately.  Upon defining 
\begin{equation}
V_{\ww}  = k \frac{(\aa - \bb)}{\aa} J_{\ww } \, , \quad V_{\wwb}  = k (\aa - \bb)   J_{\wwb } \,,\nonumber
\end{equation}
 the equations of motion can be recast as 
\begin{equation}
    \nabla_\ww  J_{\wwb } + [ J_\ww , J_{\wwb}] = 0 \, , \quad    \nabla_{\wwb}  J_{\ww  } +  \bb [ J_\ww , J_{\wwb}] = 0 \, .\nonumber
    \end{equation}
 We construct the two-dimensional Lax connection as follows.  We first build, using the IFT$_4$ objects, the $\mathbb{CP}^1$ dependent quantity
\begin{equation}
{\cal L} =  \pi^a \left(\pd_{a\dot{a}}+ C_{a \dot{a}} + B_{a\dot{a} } \right) \bar{e}^{\dot{a}}\, .\nonumber
\end{equation}
The components of a two-dimensional Lax are given by the $\dr \ww$ and $\dr \wwb$ legs of the above, or more precisely 
\begin{equation}\label{eq:lax_components_from_6d}
L_\ww  = \frac{\langle\pi \hat{\pi} \rangle}{\langle \pi \gamma \rangle \langle \hat\pi \hat\gamma \rangle }  \iota_\ww  {\cal L}  \, , \quad L_{\wwb}   = - \frac{\langle\pi \hat{\pi} \rangle}{\langle \hat \pi \gamma \rangle \langle \pi \hat\gamma \rangle }  \iota_{\wwb}   {\cal L} \,. 
\end{equation}
Evaluating the above upon the reduction ansatz \eqref{eq:RedAnsSimple} we obtain that 
\begin{equation}
L_{\wwb} = \nabla_{\wwb} + (\bb - \pp) J_{\wwb} \, , \quad J_{\wwb} = \left[  \frac{1}{\aa- \bb } U_- R^\nabla_{\wwb} +  \frac{1}{\aat - \bb } U^T_+ \widetilde{R}^\nabla_{\wwb}  \right]\,,\nonumber
\end{equation}
\begin{equation}
L_{\ww} = \nabla_{\ww} + \pp^{-1} (\bb - \pp)  J_{\ww } \, , \quad J_{\ww } = \left[  \frac{\aa}{\aa- \bb } U_+ R^\nabla_{\ww} +  \frac{\aat}{\aat - \bb } U_-^T \widetilde{R}^\nabla_{\ww}  \right]\,.\nonumber
\end{equation}
Flatness of this for all values of $\zeta$ is equivalent to 
\begin{equation}
    \nabla_\ww  J_{\wwb } + [ J_\ww , J_{\wwb}] = 0 \, , \quad    \nabla_{\wwb}  J_{\ww  } +  \bb [ J_\ww , J_{\wwb}] = 0 \, , \quad  F_{\ww \wwb }[B] = 0 \, , \nonumber
\end{equation}
which are the $g$, $\tilde{g}$ and $u'$ equations of motion, i.e. the equations of motion of all the edge modes.  Of course what this Lax does not encode directly is the on-shell solution for the gauge field $B$.  One can instead choose to eliminate $B$ entirely as it is non-propagating and replace in both the Lax and the action $B$ with its on-shell value   $B  = B[g, \tilde{g}, u] $ which is determined implicitly by the conditions 
\begin{align}
  \begin{split}
       \Big[ & k (U_{+}+         \mathrm{Ad}^{-1}_{g}  U_{-}^T) R^{\nabla}_{\ww}   -   \tilde{k} (  U_{-}^T+  \mathrm{Ad}^{-1}_{\tilde{g}} U_{+}) \tilde{R}^{\nabla}_{\ww}   \\     &\hspace{-0.4cm}+\mathtt{t}\sqrt{k\tilde{k}}(1-\mathrm{Ad}^{-1}_g)(U_{-}^T\tilde{R}^{\nabla}_{\ww})-\mathtt{t}^{-1}\sqrt{k\tilde{k}}(1-\mathrm{Ad}^{-1}_{\tilde{g}})(U_{+} R^{\nabla}_{\ww})\Big]_{\mathfrak{h}}-\nabla_{\ww}u^{\prime}=0 \, , 
    \end{split}\nonumber   \\ 
    \begin{split}
    \Big[ &-k (U_{-}+\mathrm{Ad}_g^{-1}U_{+}^T)R^{\nabla}_{\wwb}+\tilde{k}(U_{+}^T+\mathrm{Ad}_{\tilde{g}}^{-1}U_{-})\tilde{R}^{\nabla}_{\wwb}\\
    &+\mathtt{t}\sqrt{k\tilde{k}} (1-\mathrm{Ad}_{\tilde{g}}^{-1})(U_{-}R^{\nabla}_{\wwb})-\mathtt{t}^{-1}\sqrt{k\tilde{k}}(1-\mathrm{Ad}_g^{-1})(U_{+}^T\tilde{R}^{\nabla}_{\wwb})\Big]_{\mathfrak{h}}+\nabla_{\wwb}u^{\prime}=0 \, .
    \end{split}\nonumber
\end{align}
\subsubsection{Case of Enhanced symmetry}  
The structure of the operators $U_\pm$ suggests that simplifications may be obtained by defining  $\bar{g} = g \tilde{g}^{-1}$ such that   $  
U_\pm = (1 - \sigma^{\pm 1} \Ad_{\bar{g}} )^{-1}
 $.   Replacing $\tilde{g}$ with $\bar{g}$ (see Appendix \ref{appendix_EOM} for some useful details) we obtain 
\begin{align}
    J_{\wwb} = \left( \frac{1}{a- b} - \frac{\sigma^{-1}}{\tilde{a} - b}\right)   U_- R^\nabla_{\wwb} + \frac{\sigma^{-1}}{\tilde{a}-b} U_-   \bar{R}^\nabla_{\wwb },\,\nonumber\\
     J_{\ww} = \left( \frac{a}{a- b} - \frac{\tilde{a}\sigma}{\tilde{a} - b}\right)   U_+ R^\nabla_{\ww } + \frac{\tilde{a} \sigma }{\tilde{a}-b} U_+   \bar{R}^\nabla_{\ww  }\,.\nonumber
\end{align}
 This reveals the existence of a  point of enhanced symmetry \begin{equation}
 \label{eq:enhancement}     \frac{\aat}{\bb} =  \frac{\bb}{\aa }  \, ,  \quad   \sigma =  - \frac{\aa}{\bb} \, ,  
 \end{equation}
for the Lax becomes, after we  redefine the  spectral parameter $\zeta = \aa \mathfrak{z}$,   
\begin{equation} \label{eq:LaxEnhanced}
L_{\wwb} = \nabla_{\wwb}  - \frac{1+ \sigma \mathfrak{z} }{1 +\sigma }U_- \bar{R}^\nabla_{\wwb} \, \quad  
L_{\ww} = \nabla_{\ww}  - \frac{\mathfrak{z}^{-1} + \sigma   }{1 +\sigma }U_+ \bar{R}^\nabla_{\ww} \, . 
\end{equation}
The dependence on $g$ of the Lax drops out and we retain a theory only in terms of $\bar{g}$.\footnote{This has a natural interpretation in the context of 4d Chern-Simons theory defined with the meromorphic differential  
$$
\omega = \frac{\langle \pi \gamma  \rangle \langle \pi \hat{\gamma}  \rangle }{ \langle \pi \alpha \rangle \langle \pi \tilde{\alpha} \rangle  \langle \pi \beta\rangle^2   } e^0 \, .
$$
At the enhancement point the residues at $\pi = \alpha$ and  $\pi = \tilde{\alpha}$ are equal and opposite such that   the boundary condition defines an isotropic subalgebra of defect algebra. 
}

Indeed  we may recast the $\text{IFT}_2$   action in terms of the $g$ and $\bar{g}$ variables to find, in general, that 
\begin{equation}\label{eq:Action2dnewvars}
    \begin{split}
        S_{\text{IFT}_2}=-\int_{\mathbb{R}^2} \mathrm{d}\ww\wedge\mathrm{d}\wwb&\, \tr\,\Big[   c_1 R^\nabla_{\ww} U_+^T R^\nabla_{\wwb} - c_2 R^\nabla_{\ww} U_- R^\nabla_{\wwb}   + c_3 \bar{R}^\nabla_{\ww} U_+^T R^\nabla_{\wwb} -c_4 R^\nabla_{\ww}U_- \bar{R}^\nabla_{\wwb} 
       \\
        & + \frac{\tilde{k}}{2} \left( \bar{R}^\nabla_{\ww} R^\nabla_{\wwb} -\bar{R}^\nabla_{\wwb} R^\nabla_{\ww }  \right)  + \frac{\tilde{k}}{2} \bar{R}^\nabla_{\ww} (U_+^T- U_- ) \bar{R}^\nabla_{\wwb}  \Big] \\ 
     &  + \int_{\mathbb{R}^2}\frac{\tilde{k}}{2}{\cal L}_{\text{gWZ}}[\bar{g}^{-1} g , B]-\frac{k}{2}{\cal L}_{\text{gWZ}}[ g , B]+ \tr (u'  F[B])\,,\nonumber
        \end{split}
\end{equation}
where  
$$
c_1 =  \frac{k}{2}+   \frac{\tilde{k}}{2} - \sqrt{k \tilde{k}} \sigma \texttt{t} \, , \quad c_2 =  \frac{k}{2}+  \frac{\tilde{k}}{2} - \sqrt{k \tilde{k}}(\sigma \texttt{t} )^{-1}  \, , \quad c_3 = \sqrt{k \tilde{k}} \sigma \texttt{t}  -  \tilde{k} \, , \quad c_4 =  \sqrt{k \tilde{k}} (\sigma \texttt{t})^{-1} -   \tilde{k}\, . 
$$
At the point of enhanced symmetry we have that 
 $$
k =\tilde{k} \, , \quad \texttt{t}  = \frac{1}{\sigma }\,,
 $$
such that the $c_i$ all vanish. Making use of  the covariant PW identity 
\begin{equation}
{\cal L}_{\text{gWZ}}[\bar{g}^{-1} g , B] - {\cal L}_{\text{gWZ}}[  g , B] + {\cal L}_{\text{gWZ}}[ \bar{g} , B] + \tr( R^\nabla \wedge \bar{R}^\nabla) =  0 \, , 
\end{equation} 
allows us to conclude that at this parametric point the action becomes  
\begin{equation}\label{eq:Action2special}
    \begin{split}
        S_{\text{IFT}_2}=-\frac{k}{2}\int_{\mathbb{R}^2} \mathrm{d}\ww\wedge\mathrm{d}\wwb& \,   \tr\, \bar{R}^\nabla_{\ww} (U_+^T- U_- ) \bar{R}^\nabla_{\wwb}  - \int_{\mathbb{R}^2}\frac{k}{2} {\cal L}_{\text{gWZ}}[\bar{g},B] +  \tr (u'  F[B])\, .
        \end{split}
\end{equation} 
This action is the $\lambda$-model action with the $H$ subgroup gauged, but importantly with a constraint enforcing a flat connection.  This is exactly the set up for a Buscher implementation of (non-Abelian) T-duality of the $\lambda$-model.    Indeed, if we integrate by parts we may eliminate $B$ according to its equations of motion, 
\begin{align}
\tilde{k}(1 - \sigma^{-1})    U_- \bar{R}_{\wwb}^\nabla   \vert_\mathfrak{h} + \nabla_{\wwb}u'  = 0 \,  , \quad  
 \tilde{k}(1 - \sigma)    U_+\bar{R}_{\ww}^\nabla   \vert_\mathfrak{h} + \nabla_{\ww}u'  = 0\, , \nonumber
\end{align} 
to produce, upon gauge fixing, a NLSM for the $\dim \fg - \dim \fh $ non-pure-gauge degrees contained in $g$ and the $\dim \fh$ variables $u'$.   We can obtain a Lax formulation for this theory by putting the on-shell value of B in to the Lax of eq.~\eqref{eq:LaxEnhanced}.

\section{Example  and Comparison to the $\mathbb{Z}_2$ graded coset $\lambda$-deformation}\label{section:Examples_and_Comparison}

In the above we saw that as a result of the enhancement we obtain a gauging by a vectorial action of a subgroup $H$ of the $\lambda$-deformation of the $G$-WZW model.  The inclusion of the Lagrange multiplier term provides the understanding that this can yield the (non-Abelian) T-dual of the lambda model.    However if we truncate the theory by simply dispensing the Lagrange multiplier term we are left with the action
\begin{equation}
   S_{\text{IFT}_2} \vert_{u=0} = k S_{G/H~ \text{gWZW}}[\bar{g},  B] +  k   \int \dr \ww \wedge  \dr \wwb\,  \tr \bar{R}^\nabla_{\ww} U_- \bar{R}^\nabla_{\wwb} \, .\nonumber
\end{equation}
 
Evidently this truncated theory describes a deformation of the $G/H$ WZW model,  and so it is natural to ask what is the relationship of this theory to the known integrable  $\lambda$-deformation of the $G/H$ WZW model \cite{Sfetsos:2013wia,Hollowood:2014rla}.

The integrable  $\lambda$-deformation of the $G/H$ WZW model exists when $\frak{g} = \frak{g}^{(0)}  \oplus\frak{g}^{(1)}  $ admits a $\mathbb{Z}_2$ grading with respect to the grade zero subalgebra $\frak{h} = \frak{g}^{(0)}  $ and is given by a deformation of a $G/G$ model.  The deformation reduces the gauge symmetry from $G$ down to $H$. We let $A\in \frak{g}$ be a connection, and  $P^{(i)} A= A^{(i)} $ its projection into $\frak{g}^{(i)}$ and identify $B=  A^{(0)}$, and the action is given by
\begin{align}
 S_{\lambda-G/H} &= k S_{G/G~ \text{gWZW}}[\bar{g},  A]  + k (1- \lambda^{-1})  \int  \dr \ww \wedge  \dr \wwb\,  \tr  A^{(1)}_\ww     A^{(1)}_{\wwb} \, .   \nonumber
\end{align}
The components of $A$ are solved according to 
$$
A_\ww =  {\cal O}\bar{R}_\ww \, , \quad A_{\wwb} =  -{\cal O}^T\bar{L}_{\wwb}  \, , \quad  {\cal O} = [  P^{(0)} + \lambda^{-1} P^{(1)}  - \Ad_{\bar{g}}]^{-1}   \, ,
$$
and the Lax is given simply by 
$$
L^{{\lambda-G/H}}_\ww = \nabla_\ww + \frac{z}{\sqrt{\lambda}}  A^{(1)}_\ww \, , \quad   L^{{\lambda-G/H}}_{\wwb} = \nabla_{\wwb} + \frac{1}{z \sqrt{\lambda}}  A^{(1)}_{\wwb}   \, . 
$$
To make contact with the gauging produced above we may retain $B$ but eliminate out the $A^{(1)} $ components as \begin{align}
 S_{\lambda-G/H} &=    k S_{G/H~ \text{gWZW}}[\bar{g},  B]  + k   \int  \dr \ww \wedge  \dr \wwb\,  \tr  \left(  A^{(1)}_{\ww} \bar{R}^\nabla_{\wwb} - A^{(1)}_{\wwb} \bar{L}^\nabla_{\ww} - \lambda  A^{(1)}_{\ww}  U_+^{-1}    A^{(1)}_{\wwb} \right)  \nonumber \\  &=   k S_{G/H~ \text{gWZW}}[\bar{g},  B]  + k \lambda    \int  \dr \ww \wedge  \dr \wwb  \bar{R}^\nabla_{\ww } {\cal M} \bar{R}^\nabla_{\wwb} 
 \,  ,\nonumber
\end{align}
in which
\begin{equation}
{\cal M} =   \Ad_{\bar{g}}   \cdot  P^1  \cdot  [ 1- \lambda P^1 \cdot \Ad_{\bar{g}}  \cdot P^1]^{-1} \cdot P^1 \, .\nonumber
\end{equation}
Notice that as the gradation is preserved under the adjoint action of $H= \exp\frak{g}^{(0)}$ this action is $H$ gauge covariant. 

So whilst the integrable$\lambda$-deformation  does appear as a current-current (or more generally parafermionic) deformation to the gauged WZW model, it does not in general match to the truncation of our gauged model without placing further constraints.   We will now illustrate both this point and more generally our construction by means of an example. 

\subsection{\texorpdfstring{$\mathrm{SU}(2)/\mathrm{U(1)}$ gauging}{SU(2)/U(1) gauging}}

In this section we will demonstrate as a specific example the case where $G=\mathrm{SU}(2)$ and $H=\mathrm{U}(1)$. We begin by considering our 2d model \eqref{eq:Action2special} describing the non-Abelian T-dual of the $\lambda$-deformed gauged $\text{WZW}_2$. We  parametrize the group valued field $\bar{g}$ as
\begin{equation}
    \bar{g}=e^{\frac{i}{2}(\phi_1-\phi_2)\sigma^3}e^{i\omega\sigma^2}e^{\frac{i}{2}(\phi_1+\phi_2)\sigma^3},\quad \phi_i\in[0,2\pi),\, \omega\in [0,\pi] \, . \nonumber
\end{equation}
The vectorial action of $H$, $\delta    \bar{g}  = \epsilon [T^3 ,     \bar{g}] $, leaves invariant $\phi_1$ and $\omega$   whilst acting by shifts of $\phi_2$ which we fix with    $\phi_2=0$ (and then rename $\phi_1 = \phi)$.  
For the $\mathfrak{u}(1)$ valued field we chose the $T^3$ direction\footnote{we pick the $\mathfrak{su}(2)$ generators to be $T^j=\frac{i}{\sqrt{2}}\sigma^j$, $j\in\{1,2,3\}$.} in $\mathfrak{su}(2)$    i.e. 
\begin{equation}
    B =\left[b_{\ww} (\ww,\wwb)\mathrm{d}\ww+b_{\wwb} (\ww,\wwb)\mathrm{d}\wwb\right] T^3\,.\nonumber
\end{equation} 
We can eliminate $B$ from the action  eq.~\eqref{eq:Action2special} via its   on-shell solution
\begin{equation}
\begin{aligned}
b_{\ww}(\ww,\wwb) & =   \frac{1}{\sqrt{2} } \frac{1}{\sigma^2 -1 } \left( f(\phi) \cot^2 \omega  \partial_{\ww} \chi + 2\sigma  \sin 2 \phi   \cot \omega \partial_{\ww} \omega  - (1+ \sigma)^2 \partial_\ww u \right), \\
b_{\wwb}(\ww,\wwb) & =   -\frac{1}{\sqrt{2} } \frac{  1}{\sigma^2 -1 } \left( f(\phi)  \cot^2 \omega  \partial_{\wwb} \chi + 2\sigma  \sin 2 \phi   \cot \omega \partial_{\wwb} \omega  - (1+ \sigma)^2 \partial_{\wwb} u \right) \, , 
\end{aligned}\nonumber
\end{equation}
in which $\chi = \phi - u $ and $f(\phi)=\left[1+\sigma^2-2\sigma\cos(2\phi) \right]$.  Having done so,   the action \eqref{eq:Action2special} becomes that of a metric only non-linear sigma model with target space 
\begin{equation} \label{sigmamodel}
    \frac{1}{k}\mathrm{d}s^2 =   \frac{1-\sigma}{1+\sigma}\left( \cot^2\omega  \dr \chi^2+\mathrm{d}\omega^2\right)+\frac{4\sigma}{1-\sigma^2}\left(\cot\omega\sin\phi \mathrm{d}\chi +\cos\phi \mathrm{d}\omega\right)^2  + \frac{1+\sigma}{1-\sigma}  \dr u^2 \, . 
    \end{equation} 

Using equations \eqref{eq:lax_components_from_6d} and \eqref{eq:LaxEnhanced} we can write a Lax connection for this sigma model of the form  $ L =\mathrm{d} + B   + l   $ 
 with components  $ l_\ww  = l_\ww^A T_A $ given by
\begin{equation}
\begin{aligned} \label{eq:Lax_components_uneq0}
(l_\ww )^A  =  \sqrt{2}   c_\ww(\mathfrak{z})   \left\{ \frac{  \cos\phi \cot\omega \pd_\ww \chi - \sin\phi \pd_\ww \omega  \  }{ 1+\sigma }  \, , \frac{ \sin\phi \cot\omega \pd_\ww \chi + \cos\phi \pd_\ww \omega   }{\sigma-1 }   \,,  \frac{\pd_{\ww} u}{\sigma-1} \right\}\, , \\
(l_{\wwb} )^A  =  \sqrt{2}    c_{\wwb} (\mathfrak{z})  \left\{\frac{  \cos\phi \cot\omega \pd_{\wwb} \chi - \sin\phi \pd_{\wwb}  \omega  \  }{ 1+\sigma  }  \, , \frac{ \sin\phi \cot\omega \pd_{\wwb} \chi + \cos\phi \pd_{\wwb} \omega   }{ 1 - \sigma }   \,,  \frac{\pd_{\wwb} u}{ 1  -\sigma}  \right\} \, ,
\end{aligned} 
\end{equation} 
where the dependence on the spectral parameter, $\mathfrak{z}$, is encoded in the functions
\begin{equation} \label{eq:cs} c_\ww(\mathfrak{z}) = \frac{1+ \mathfrak{z} \sigma}{\mathfrak{z}(1+\sigma) } \,  , \quad  c_{\wwb} (\mathfrak{z}) = \sigma \frac{1+ \mathfrak{z}\sigma}{ (1+\sigma) } \,  . 
\end{equation}

The flatness of this Lax is equivalent to the equations of motion of \eqref{sigmamodel} which we write here for completeness:
\begin{equation} 
    \begin{split}
        \delta u: \,\, & 2\sigma \cot\omega \partial_{\wwb}\chi \left[ \cot\omega \sin(2\phi) \partial_{\ww}\chi +\cos(2\phi)\partial_{\ww}\omega  \right] \\
        &+2\sigma \partial_{\wwb}\omega \left[ \cos(2\phi) \cot\omega \partial_{\ww}\chi - \sin(2\phi) \partial_{\ww}\omega  \right] -(1+\sigma )^2 \partial_{\ww}\partial_{\wwb}u=0 \,,
    \end{split}\nonumber
\end{equation}

\begin{equation}
    \begin{split}
        \delta \phi:\,\, & 2\sigma \cos(2\phi)(\partial_{\wwb}\omega\partial_{\ww}u + \partial_{\wwb}u\partial_{\ww}\omega)- f(\phi)  \csc^2\omega (\partial_{\wwb}\omega\partial_{\ww}\chi + \partial_{\wwb}\chi\partial_{\ww}\omega)\\
        &+2\sigma \cot\omega\sin(2\phi)\left[ \partial_{\wwb}u\partial_{\ww}\chi + \partial_{\wwb}\chi (\partial_{\ww}u+\partial_{\ww}\chi)-\partial_{\wwb}\omega\partial_{\ww}\omega \right]+2\sigma\sin(2\phi)\partial_{\ww}\partial_{\wwb}\omega\\
&        +f(\phi) \cot\omega \partial_{\ww}\partial_{\wwb}\chi=0\,,
    \end{split}\nonumber
\end{equation}

\begin{equation}
    \begin{split}
        \delta \omega: \,\, & -2\sigma\cos(2\phi)\cot\omega \partial_{\wwb}u\partial_{\ww}\chi+2\sigma \sin(2\phi)\left[ \partial_{\wwb}\omega (\partial_{\ww}u+\partial_{\ww}\chi )+\partial_{\wwb}u\partial_{\ww}\omega - \cot\omega \partial_{\ww}\partial_{\wwb}\chi \right]\\
        &+\partial_{\wwb}\chi \Big( -2\sigma \cos(2\phi)\cot\omega \partial_{\ww}u - \left[ 1+\sigma^2-2\sigma \cos(2\phi)\cos(2\omega) \right] \cot\omega \csc^2\omega \partial_{\ww}\chi\Big)\\
        &-\left[ 1+\sigma^2+2\sigma \cos(2\phi) \right]\partial_{\ww}\partial_{\wwb}\omega =0\,.
    \end{split}\nonumber
\end{equation}

We now perform a truncation of the above system by setting $u= 0$. At the level of the target space metric we immediately obtain 

\begin{equation}\label{eq:coset_metric}
   \frac{1}{k} \mathrm{d}s^2\Big|_{u=0}= \left[\frac{1-\sigma}{1+\sigma}\left( \cot^2\omega\mathrm{d}\phi^2+\mathrm{d}\omega^2\right)+\frac{4\sigma}{1-\sigma^2}\left(\cot\omega\sin\phi \mathrm{d}\phi+\cos\phi\mathrm{d}\omega\right)^2\right]\,.
\end{equation} 
which we recognise as the $\lambda$-model, where the eponymous parameter is identified with $\sigma$, see agreement with equation (5.2) of \cite{Sfetsos:2014cea} as well as the results in \cite{Sfetsos:2013wia}. This is in contrast to the general discussion in the beginning of this section, where we point out the truncated theory would not in general be the $\lambda$-model, but in this specific example we find that it is.

One might at first anticipate that the truncation $u=0$ applied to \eqref{eq:Lax_components_uneq0} would also yield a Lax for the truncated theory.   This however is not true; by simply substituting $u=0$ and then calculating the flatness of the Lax one finds a disagreement with the equations of motion that would be derived from the truncated theory \eqref{eq:coset_metric} - the reason is that the Lax encodes a first order constraint that $F[B] =0 $ and this constraint remains enforced even after having truncated the Lax. 

We can, however, recognise that the true Lax connection of the non-linear sigma model \eqref{eq:coset_metric} is structurally similar to the $u=0$ limit of \eqref{eq:Lax_components_uneq0}; the difference lies in the functional dependence on the spectral parameter, which  instead of   eq.~\eqref{eq:cs}, is required to be of the form,
$ c_{\ww}(z) =z $ and $   c_{\wwb}(z) = z^{-1} $.  
There is no redefinition of $
\mathfrak{z} = f(z) $ such that the $c(\mathfrak{z})$ of eq.~\eqref{eq:cs}  can be made to agree with the   $c(z)$  here. Indeed, the critical feature is that in the commutator term of the Lax field strength these factors conspired to give an ${\cal O}(z^0)$ contribution that sources a non-zero result for $F[B]$.\footnote{It is exactly this mechanism that requires the $\mathbb{Z}_2$ grading; since the $l_\ww$  and $ l_{\wwb}$ lie in $\mathfrak{g}^{(1)}$  the  ${\cal O}(z^0)$   contribution from $[l_\ww, l_{\wwb}]$ lies only in $\mathfrak{g}^{(0)}$ and combines with $F[B]$ rather than presenting further independent equations. }  \\

In summary, this is a particular instance where we have an ``accidental" agreement at the level of the action between the  truncated non-Abelian T-dual theory and the coset $\lambda$-deformed gauged WZW model, however this does not directly extend to the Lax formulation.

\section{Conclusions}

The aim of this work has been to clarify the higher dimensional origin of gauged integrable models. The key concept presented here is a refined framework in which one can formulate gauged holomorphic Chern-Simons theories on twistor space, based on the diamond construction of \cite{Bittleston:2020hfv,Cole:2023umd,Cole:2024sje} in a conceptually clear way. It is from these theories that we can then extract new four and two-dimensional integrable models, as many works have recently demonstrated in the literature. The theories in six dimensions consist of two Lie algebra valued connections, ${\cal A}\in \Omega^1(\mathbb{PT})\otimes\mathfrak{g}$ and ${\cal B}\in \Omega^1(\mathbb{PT})\otimes \mathfrak{h}$, as well as an auxiliary meromorphic three-form $\Omega\in \Omega^{(3,0)}(\mathbb{PT})$ whose holomorphic data, together with the reduction data, completely determines the lower dimensional theories. By choosing to describe the theories in terms of the fields ${\cal C}={\cal A}-{\cal B}$ and ${\cal B}$ instead, and by using the language of Cartan connections, we acquire a manifestly covariant formulation which motivates the need for a boundary term in the action, previously included in an \textit{ad hoc} fashion. As a demonstration we use the holomorphic data corresponding to the $\lambda$-model in the ungauged case, i.e. a (1,1,2) pole structure for $\Omega$. Localizing the theory on the poles in the $\mathbb{CP}^1$ fibre we obtain new families of four-dimensional integrable field theories described by the action \eqref{4d_action}, whose equations of motion correspond to the anti-self duality of a four-dimensional connection $A$. Upon dimensional reduction, we find new two-dimensional integrable models given by \eqref{eq:Action2d} containing covariant gauged currents from the edge modes of each single pole, with interactions between them. Common to the un-gauged discussion of \cite{Cole:2023umd} there is a parametric point for which the theories gain additional local symmetries such that  only one out of the two edge mode fields contributes to the dynamics. 

In this way we obtain a  two-dimensional model that corresponds to the gauging of the  $\lambda$-deformed $\mathrm{WZW}_2$ by means of a flat connection.   This can be understood as implementing an non-Abelian T-dualisation of the $\lambda$-deformed $\mathrm{WZW}_2$.    Indeed the emergence of a two-dimensional Lagrange multiplier term can be directly traced to a coupling ${\cal C}\wedge {\cal F}[{\cal B}] $ that is required for gauge invariance in six-dimensions.

In the undeformed $\lambda\rightarrow 0 $ considered in \cite{Cole:2024sje} it is possible to dispense of the Lagrange multiplier term all-together and obtain the $G/H$ coset CFT realised as gauged WZW.   Here however, once the deformation is active, $\lambda\neq 0$, dropping the requirement that the gauging is flat is no longer {\em in general} something that can be done in a way that preserves integrability and we contrast this truncation with the known integrable $\lambda$-deformation of the $G/H$ CFT.  

 As an exception however to this general picture  we study the case where $G/H=\mathrm{SU}(2)/\mathrm{U}(1)$. In this specific case, one {\it can} achieve matching with the $\lambda$-model after truncating the theory at the level of the action by discarding the term enforcing the constraint. Nevertheless, the truncated Lax connection still encodes information from the constraint and thus needs further modification to exactly match the one for the $\lambda$-deformed $\mathrm{WZW}_2$.

 \subsection{Future Directions}

The six-dimensional gauging of the diamond has thus provided a concrete realization of dualisation in integrable models, opening several avenues for further study:

\begin{itemize}
    \item \textbf{Poisson–Lie Dualisation} The current construction produces non-Abelian dualisations. Given the close connection between integrable models and Poisson–Lie duality \cite{Klimcik:1995ux}, it is natural to ask whether a six-dimensional origin for general Poisson–Lie dualisations can be formulated, especially since these do not admit a straightforward Buscher-type \cite{Buscher:1987sk} implementation.  
    
\item \textbf{Fermionisation/Bosonisation} Could other dualities, such as fermionisation or bosonisation, be understood from a higher-dimensional perspective? These are known to admit gauging procedures in two dimensions \cite{Burgess:1994np}, suggesting a potential 6d origin.

\item \textbf{Quantum Aspects of Non-Abelian Dualisation} In the Abelian case, ungauged and gauged models are quantum mechanically equivalent. For non-Abelian dualisations, the quantum status is less clear. Extending the current framework to address arbitrary genus 2d worldsheets and embedding them into the diamond remains an important challenge, with recent work \cite{Jarov:2025bmq} in this direction.   At the 6d level even in the case of Abelian gaugings it is not clear that gauged models and ungauged models are equivalent and elucidating this both classically and quantum mechanically may be profitable.

\item \textbf{Boundary Conditions and Defects}  A general expectation is that some dualisations could correspond to different choices of boundary conditions. This is motivated by the way that pairs of dual sigma models correspond to complementary lagrangian subalgebras of a Drinfeld double.  It remains somewhat open to address the full classification of boundary conditions in 6d, and to identify the cases for which the 2d theories are a classical/quantum dual pair.  Here   we obtained a dualisation through our 6d gauging construction; one might consider if it is possible to  interpret gauged 6d models as ungauged models with alternative boundary conditions.

\item \textbf{Connections to Lattice Models} Recent advances  \cite{Appadu:2017fff,Appadu:2018ioy,Ashwinkumar:2023zbu} suggest that $\lambda$-models and more generally 4d Chern–Simons theory setups can provide a pathway to light-cone lattice quantisation and associated spin-chain descriptions.  At the same time there have been exciting advances in framing dualisation in lattice systems \cite{Lootens:2023wnl,Vancraeynest-DeCuiper:2025fpn},  and it would be valuable to relate the continuum dualisation mechanisms developed here to these lattice constructions.
 
\end{itemize}

Finally, it is worth noting that one original motivation of this study was to identify a six-dimensional origin of the integrable $\lambda$-deformation of the $ 
G/H$  gauged WZW model. Our results show that this is {\it not} directly realized within the current framework. Achieving this goal likely requires explicitly incorporating the $\mathbb{Z}_2$ (or $\mathbb{Z}_4$) grading of  the undeformed integrable structure, potentially via a modified gauging procedure, alternative boundary conditions, or a yet-to-be-determined construction (see \cite{Cole:2024skp}, \cite{Berkovits:2024reg} for cases in which such grading is crucial and also the more general branch cut defects of \cite{Costello:2019tri}). 

Beyond the specific gauging mechanism analysed here, there remain several structural questions about the holomorphic Chern–Simons origin of integrable models. A natural next step is a systematic study of admissible boundary conditions for the six-dimensional fields on twistor space and the corresponding web of dual two-dimensional theories they generate. More generally, understanding how different meromorphic structures for 
 $\Omega$ organise the full landscape of integrable deformations, how quantum considerations enter this picture, and whether any of these constructions admit a string-theoretic or gravitational interpretation, will be key arenas for development. These questions sit somewhat beyond the scope of the present work, but they form a broader context in which the gauging of the diamond developed here can be viewed as one concrete piece.

\section*{Acknowledgments}
This work is supported by the STFC consolidated grant ST/X000648/1.  DC is supported by an STFC studentship and JM supported in part by EPSRC grant EP/W524694/1 and by an STFC studentship.  We thank Lewis Cole for useful discussions, and the organisers of the \href{https://indico.fysik.su.se/event/8807/}{Integrability Dualities and Deformations 2025} conference during which some of this work was developed and presented. 

\appendix 
\section{Calculation of equations of motion details}\label{appendix_EOM}
In this section we note a number of calculation details that are used in the derivation of the four-dimensional action and its equations of motion.   Although  algebraic in nature, these details are sufficiently tedious to reproduce that it warrants recording them for posterity.  

For the gauge transformation
$$
B \mapsto   h^{-1} B h + h^{-1} \dr h \, , \quad g\mapsto h^{-1} g h \, , 
$$
we have the covariant Maurer-Cartan forms
$$
R^\nabla = \dr g g^{-1} + B - g B g^{-1} \mapsto h^{-1} R^\nabla  h \, , 
$$
which obey the covariant Maurer-Cartan identity 
$$
\nabla R^\nabla \equiv  \dr R^\nabla + B \wedge R^\nabla  + R^\nabla \wedge B  =R^\nabla\wedge R^\nabla  + F[B] - g  F[B]  g^{-1} \, .  
$$
Useful identities to handle the variation of these covariant Maurer-Cartan forms acting on some object $X$ are 
\begin{align*}
   & \tr\left[ ( \delta_g R_\mu^\nabla) X \right] =  \tr[ \delta g g^{-1} \left( -\nabla_\mu X  + [R^\nabla_\mu , X] \right) ] + \text{total derivative} \, ,  \\ 
& \tr \left[   \left(\delta_B R^{\nabla}_{\mu}\right)X \right]= \tr \left[(1-\mathrm{Ad}^{-1}_g)X\, \delta B_{\mu}\right]    \, .  
\end{align*}

Under arbitrary variations of $g$ and $B$ the gauge invariant WZ varies   
\begin{align*}
&   \delta_g {\cal L}_{\text{gWZ}}[g,B]  = \tr \left[  \delta g g^{-1} \left(\nabla R^\nabla  - 2 F[B]\right) \right]
  + \text{total derivative} \,  , \\ 
&\delta_B {\cal L}_{\text{gWZ}}[g,B] = - \tr\left[  \delta B \wedge (R^\nabla +  L^\nabla) \right]  \, . 
\end{align*}

The components of the 4d gauge field are given by 
$$
C_{\usf}  = - U_+ R^\nabla_\wsf - U_-^T \tilde{R}^\nabla_\zsf \, , \quad C_{\husf} =  - U_- R^\nabla_{\hwsf} - U_+^T \tilde{R}^\nabla_{\hzsf}\,,
$$ 
where 
\begin{equation}\label{U_matrices}
U_\pm = \left( 1 - \sigma^{\pm 1 } \Ad_g\circ \Ad_{\tilde{g}}^{-1} \right)^{-1}  \, , \quad U_\pm + U_\mp^T = \textrm{id} \, .
\end{equation}
Using  
\begin{align*}
&\delta_g U_\pm(X) =   U_\pm \left(  \delta_g X  -   [\delta g g^{-1} , U_\mp^T (X) ] \right) =U_\pm \left(  \delta_g X  -   [\delta g g^{-1} ,X  ] +  [\delta g g^{-1} , U_\pm(X) ] \right)  \, ,\\
  &\delta_{\tilde g}U_{\pm}(X)=U_{\pm}\delta_{\tilde g}X+U_{\mp}^{T}[\delta\tilde{g} \tilde{g}^{-1},U_{\pm}(X)] \, ,\end{align*}
we have 
\begin{align*}
\delta_g C_{\usf} &= U_+ \left( -\delta_g R^\nabla_{\wsf} + [\delta g g^{-1},  R^\nabla_{\wsf} +C_{\usf}    ] \right) \,, \\
\delta_g C_{\husf} &= U_- \left( -\delta_g R^\nabla_{\hwsf} + [\delta g g^{-1},  R^\nabla_{\hwsf} +C_{\husf}    ] \right) \, . 
\end{align*}
Let us define 
$$
{\cal L}_0 = \tr \left(  C_\usf ( R^\nabla_{\hwsf}  -\tilde{R}^\nabla_{\hzsf}) -  C_{\husf}   ( R^\nabla_{\wsf}  -\tilde{R}^\nabla_{\zsf})   \right) \, ,
$$
whose variation under $g$ reads
\begin{equation*}
\begin{split}
  \delta_g {\cal L}_0 & = \tr \left[ \delta_g   C_\usf ( R^\nabla_{\hwsf}  -\tilde{R}^\nabla_{\hzsf}) -  \delta_g  C_{\husf}   ( R^\nabla_{\wsf}  -\tilde{R}^\nabla_{\zsf})  + C_\usf  \delta_g R^\nabla_{\hwsf} - C_{\husf}   \delta_g    R^\nabla_{\wsf} \right]\\ 
& = \tr \left( U_+^{-1} \delta_g   C_\usf ( C_{\husf} +  R^\nabla_{\hwsf} ) - U_-^{-1} \delta_g  C_{\husf}   (C_\usf  +  R^\nabla_{\wsf} )  + C_\usf  \delta_g R^\nabla_{\hwsf} - C_{\husf}   \delta_g    R^\nabla_{\wsf} \right) \\
&=  \tr \left[ \delta g   g^{-1} \left( \nabla_{\wsf} R^\nabla_{\hwsf} -\nabla_{\hwsf} R^\nabla_{\wsf}  + 2 \nabla_{\wsf} C_{\husf}- 2 \nabla_{\hwsf} C_{\usf}  + 2 [ C_{\usf}, C_{\husf} ]\right) \right] \\
&=   \iota_{\hwsf} \iota_{\wsf} \tr \left[ \delta g   g^{-1}  \left( \nabla R^\nabla + 2 \nabla C + 2 C \wedge C \right) \right]\,.
\end{split}
\end{equation*}
We then consider 
$$
{\cal L}^{tot} = {\cal L}_0[g,\tilde{g} , B ] + \iota_{\hwsf} \iota_{\wsf} {\cal L}_{\text{gWZ}}[g,B] -  \iota_{\hzsf} \iota_{\zsf} {\cal L}_{\text{gWZ}}[\tilde{g},B] - 2  \iota_{\hvsf}\iota_{\vsf}  \tr \left( u F[B] \right)\,.
$$ 
Denoting $C= A- B$, $A^g = g^{-1} A g + g^{-1} \mathrm{d} g$,   $ 
Q= A - A^g$ and $ \tilde{Q} = A - A^{\tilde{g}} $  
we  have the final result for the variation of the four dimensional action
\begin{align*}
\label{eq:variationLtot}
    \delta {\cal L}^{tot} =& -2 \iota_{\hwsf} \iota_{\wsf} \tr\left( \delta g g^{-1}   F[A]  -  \delta B\wedge Q \right) +2 \iota_{\hzsf}\iota_{\zsf}  \tr \left(\delta \tilde{g}  \tilde{g}^{-1}   F[A]  -  \delta B\wedge \tilde{Q} \right) \nonumber \\
    & - 2 \iota_{\hvsf} \iota_{\vsf}  \tr \left( \delta u  F[B] \right) - 2  \iota_{\hvsf}\iota_{\vsf}  \tr \left( \delta B\wedge \nabla u  \right)  \, . 
\end{align*}

In investigating the point of enhancement, eq.~\eqref{eq:enhancement}, we find it useful to eliminate $\tilde{g}$ by means of the change of variables  $\bar{g} = g \tilde{g}^{-1}$ for which we have 

$$\bar{R} = \dr (g \tilde{g}^{-1} ) \tilde{g} g^{-1} = R - 
\Ad_{\bar{g}} \tilde{R}\,, $$
  and also

$$
\tilde{R}^\nabla  = \Ad_{\bar{g}}^{-1} ( -\bar{R}^\nabla  +  R^\nabla  )\, .
$$
The change of variables is motivated by the form of the $U_\pm$ operators $U_\pm = (1 - \sigma^{\pm 1} \Ad_{\bar{g}} )^{-1}$, and useful identities are 
$$
U_\pm^T  \Ad_{\bar{g}}^{-1} = - \sigma^{\mp 1} U_\mp \, . 
$$
In these variables the components of the 4d gauge field are 
\begin{align*}
C_{\usf}  &= - U_+ R^\nabla_\wsf - U_-^T \tilde{R}^\nabla_\zsf  =  - U_+ \left[ R^\nabla_\wsf -  \sigma   R^\nabla_\zsf + \bar{R}^\nabla_\zsf \right] \, ,  \\  C_{\husf} &=  - U_- R^\nabla_{\hwsf} - U_+^T \tilde{R}^\nabla_{\hzsf} =  - U_- \left[ R^\nabla_{\hwsf} -  \sigma^{-1}   R^\nabla_{\hzsf} + \bar{R}^\nabla_{\hzsf} \right] \, .     
\end{align*}

\section{Full expression for the four-dimensional action}\label{appendix_4d_action}

In the following, we provide more details on the four-dimensional action \eqref{4d_action} in the coordinates defined by \eqref{eq:newcoordinates}. Suppose we have a one-form $X = X_{a \dot{a}} \dr x^{a\dot{a}} = X_\zsf \dr \zsf +X_{\hzsf} \dr \hzsf + X_\wsf \dr \wsf +X_{\hwsf} \dr \hwsf    $, then the coordinate transformation rules are that

\begin{equation*}
 \begin{aligned}
   \langle \alpha \beta \rangle  X_\wsf &=    \langle \alpha \gamma \rangle  X_\ww +   \langle \alpha \hat\gamma \rangle   X_{\zzb} \, , \quad  
    \langle \alpha \beta \rangle  X_{\hwsf}  =    \langle \alpha \hat\gamma \rangle   X_{\wwb}  -  \langle \alpha \gamma \rangle X_\zz  \, ,  \\  
       \langle \tilde \alpha \beta \rangle  X_\zsf &=    \langle  \tilde  \alpha \gamma \rangle  X_\ww +   \langle  \tilde  \alpha \hat\gamma \rangle   X_{\zzb}  \, , \quad 
    \langle \tilde  \alpha \beta \rangle  X_{\hzsf} =    \langle \tilde  \alpha \hat\gamma \rangle   X_{\wwb}  -  \langle \tilde  \alpha \gamma \rangle X_\zz   \, . 
    \end{aligned}
\end{equation*}
Manipulation of the action proceeds with the useful identity for the wedge product of a two-form $q$ with the self-dual  $\Sigma_\pi = \pi_a \pi_b \epsilon_{\dot{a} \dot{b} } \dr x^{a \dot{a}} \wedge \dr x^{b \dot{b}}$  
\begin{equation*}
    \Sigma_\pi \wedge q = 2\,\mathrm{vol}_4 \left(   q_{\zzb \wwb } \langle \pi \hat\gamma \rangle^2 +q_{\zz \ww } \langle \pi  \gamma \rangle^2  + \left( q_{\zz \zzb } + q_{\ww \wwb} \right)   \langle \pi  \gamma \rangle \langle \pi \hat\gamma \rangle   \right) \,,
\end{equation*}
and related  special cases 
\begin{align*} 
 \iota_{\hzsf }\iota_\zsf q  = \frac{1}{\langle \tilde\alpha \beta\rangle^2 } \left( \langle\tilde{\alpha} \hat{\gamma }\rangle^2 q_{\zzb \wwb } + \langle\tilde{\alpha}  \gamma \rangle^2 q_{\zz  \ww  }  + \langle\tilde{\alpha}  \gamma \rangle\langle\tilde{\alpha} \hat{\gamma }\rangle   (q_{\ww \wwb} +q_{\zz \zzb })    \right) \, , \\
  \iota_{\hwsf }\iota_\wsf q= \frac{1}{\langle  \alpha \beta\rangle^2 } \left( \langle  \alpha  \hat{\gamma }\rangle^2 q_{\zzb \wwb } + \langle \alpha   \gamma \rangle^2 q_{\zz  \ww  }  + \langle \alpha   \gamma \rangle\langle \alpha \hat{\gamma }\rangle   (q_{\ww \wwb} +q_{\zz \zzb })    \right) \, ,  \\
      \mathrm{d}\usf\wedge\mathrm{d}\husf=-\langle\beta\gamma\rangle\langle\beta\hat\gamma\rangle\,\mathrm{d}\ww\wedge\mathrm{d}\wwb-\langle\beta\hat\gamma\rangle^2\mathrm{d}\ww\wedge\mathrm{d}\zz-\langle\beta\gamma\rangle \langle\beta\hat\gamma\rangle\, \mathrm{d}\zz\wedge\mathrm{d}\zzb+\langle\beta\gamma\rangle^2\mathrm{d}\wwb\wedge\mathrm{d}\zzb\,.
\end{align*} 

We proceed by applying this change of basis and write the four-dimensional action \eqref{4d_action} in holomorphic coordinates. Due to the long nature of the expression we choose to break the action into terms containing couplings of the the form $R^{\nabla}R^{\nabla}$ (and similarly for the currents of the field $\tilde{g}$), terms that couple the currents of fields from both single order poles, $R^{\nabla}\tilde{R}^{\nabla}$, as well as contributions from the gauged WZ Lagrangians 
and the Lagrange multiplier term for $u$. The full expression reads:

\begin{equation}\label{4d_action_in_new_coords}
    \begin{split}
        S_{\mathrm{IFT}_4}&=\frac{K}{\langle\alpha\tilde\alpha\rangle}\int_{\mathbb{R}^4}\mathrm{vol}_4\,\tr \Big( \mathcal{Y}_1^{R^{\nabla}R^{\nabla}}+\mathcal{Y}_2^{\tilde{R}^{\nabla}\tilde{R}^{\nabla}}+{\cal Y}_3^{R^{\nabla}\tilde{R}^{\nabla}}+\mathcal{Y}_4^{\text{gWZ}}\Big)+S^{(u)}\,,
    \end{split}
\end{equation}
where
\begin{equation} 
\begin{split}
   {\cal Y}_1^{R^{\nabla}R^{\nabla}}=-\frac{1}{2\langle\alpha\beta\rangle^2}&\Big[\langle\alpha\gamma\rangle\langle\alpha\hat\gamma\rangle\Big(R^{\nabla}_{\ww}(U_{+}^{T}-U_{-})R^{\nabla}_{\wwb}-R^{\nabla}_{\zzb}(U_{+}^{T}-U_{-})R^{\nabla}_{\zz}\Big)\\
        &-\langle\alpha\gamma\rangle^2 R^{\nabla}_{\ww}(U_{+}^{T}-U_{-})R^{\nabla}_{\zz}+\langle\alpha\hat\gamma\rangle^2R^{\nabla}_{\zzb}(U_{+}^{T}-U_{-})R^{\nabla}_{\wwb}\Big]\,,
        \end{split}\nonumber
\end{equation}
and for the contributions of the same kind from the pole $\pi=\tilde{\alpha}$, the expression is the same as the above but with the pole data exchanged, i.e.

\begin{equation}
    {\cal Y}_2^{\tilde{R}^{\nabla}\tilde{R}^{\nabla}}={\cal Y}_1^{R^{\nabla}R^{\nabla}}\quad \text{with}\quad(\alpha\longleftrightarrow\tilde{\alpha}\quad, \quad R^{\nabla}\longleftrightarrow \tilde{R}^{\nabla})\,.\nonumber
\end{equation}
Then the contribution from mixed couplings is
\begin{equation}
\begin{split}
    {\cal Y}_3^{R^{\nabla}\tilde{R}^{\nabla}}&=\frac{1}{\langle\alpha\beta\rangle\langle\tilde\alpha\beta\rangle} \Big[ \langle\alpha\gamma\rangle\langle\tilde\alpha\hat\gamma\rangle R^{\nabla}_{\ww}U_{+}^{T}\tilde{R}^{\nabla}_{\wwb}-\langle\alpha\gamma\rangle\langle\tilde\alpha\gamma\rangle R^{\nabla}_{\ww}U_{+}^{T}\tilde{R}^{\nabla}_{\zz} \\
        & +\langle\alpha\hat\gamma\rangle \langle\tilde\alpha\hat\gamma\rangle R^{\nabla}_{\zzb}U_{+}^{T}\tilde{R}^{\nabla}_{\wwb}-\langle\alpha\hat\gamma\rangle \langle\tilde\alpha\gamma\rangle R^{\nabla}_{\zzb}U_{+}^{T} \tilde{R}^{\nabla}_{\zz}\Big]\\
          &-\frac{1}{\langle\tilde\alpha\beta\rangle\langle\alpha\beta\rangle} \Big[ \langle\tilde\alpha\gamma\rangle\langle\alpha\hat\gamma\rangle \tilde{R}^{\nabla}_{\ww}U_{-}R^{\nabla}_{\wwb}-\langle\tilde\alpha\gamma\rangle\langle\alpha\gamma\rangle \tilde{R}^{\nabla}_{\ww}U_{-}R^{\nabla}_{\zz} \\
        & +\langle\tilde\alpha\hat\gamma\rangle \langle\alpha\hat\gamma\rangle \tilde{R}^{\nabla}_{\zzb}U_{-}R^{\nabla}_{\wwb}-\langle\tilde\alpha\hat\gamma\rangle \langle\alpha\gamma\rangle \tilde{R}^{\nabla}_{\zzb}U_{-} R^{\nabla}_{\zz}\Big]\,,
    \end{split}\nonumber
\end{equation}
and the gauged WZ terms

\begin{equation}
    \begin{split}
        {\cal Y}_4^{\text{gWZ}}&=-\frac{1}{2\langle \alpha \beta\rangle^2}\Big[\langle\alpha\hat\gamma\rangle^2\iota_{\wwb}\iota_{\zzb}{\cal L}_{\text{gWZ}}[g,B]+\langle \alpha\gamma\rangle^2\iota_{\ww}\iota_{\zz}{\cal L}_{\text{gWZ}}[g,B]\\
        &+\langle \alpha\gamma\rangle\langle \alpha\hat\gamma\rangle\Big(\iota_{\wwb}\iota_{\ww}{\cal L}_{\text{gWZ}}[g,B]+\iota_{\zzb}\iota_{\zz}{\cal L}_{\text{gWZ}}[g,B] \Big)\Big]\\ 
        &+\frac{1}{2\langle \tilde\alpha \beta\rangle^2}\Big[\langle \tilde\alpha\hat\gamma\rangle^2\iota_{\wwb}\iota_{\zzb}{\cal L}_{\text{gWZ}}[ \tilde g,B]+\langle  \tilde \alpha\gamma\rangle^2\iota_{\ww}\iota_{\zz}{\cal L}_{\text{gWZ}}[ \tilde g,B]\\
        &+\langle  \tilde \alpha\gamma\rangle\langle  \tilde \alpha\hat\gamma\rangle\Big(\iota_{\wwb}\iota_{\ww}{\cal L}_{\text{gWZ}}[ \tilde g,B]+\iota_{\zzb}\iota_{\zz}{\cal L}_{\text{gWZ}}[ \tilde g,B] \Big)\Big]\,,
    \end{split}\nonumber
\end{equation}
while

\begin{equation}
    \begin{split}
        S^{(u)}= -\frac{K}{\langle\alpha\tilde\alpha\rangle}\int_{\mathbb{R}^4}\tr (u F[B])\wedge&\Big[-\langle\beta\gamma\rangle\langle\beta\hat\gamma\rangle\,\mathrm{d}\ww\wedge\mathrm{d}\wwb-\langle\beta\hat\gamma\rangle^2\mathrm{d}\ww\wedge\mathrm{d}\zz\\
        &-\langle\beta\gamma\rangle \langle\beta\hat\gamma\rangle\, \mathrm{d}\zz\wedge\mathrm{d}\zzb+\langle\beta\gamma\rangle^2\mathrm{d}\wwb\wedge\mathrm{d}\zzb\Big]\,.
    \end{split}\nonumber
\end{equation}

\section{2d Equations of Motion }\label{sec:2deqm}
Here we derive the $g$ equations of motion for the two-dimensional action \eqref{eq:Action2d} (the $\tilde{g}$ equations can be found similarly). The $g$-dependent integrand is comprised of the following terms, using the identity \eqref{U_matrices} we have

\begin{align*}
    \textrm{term}_1 &=  - k\, \tr\,R^\nabla_\ww U_- R^\nabla_{\wwb} ,\\ 
    \textrm{term}_2 &= \frac{k}{2} \tr\, R^\nabla_\ww R^\nabla_{\wwb} +\frac{k}{2} \iota_{\wwb}\iota_{\ww}{\cal L}_{\text{gWZ}}[g,B], \\
    \textrm{term}_3 &= \tilde{k} \, \tr\,\widetilde{R}^\nabla_{\ww } U^T_+  \widetilde{R}^\nabla_{\wwb } , \\
     \textrm{term}_4 &= \sqrt{k \tilde{k}} \texttt{t}\, \tr\, \widetilde{R}^\nabla_\ww U_- R^\nabla_{\wwb}  , \\ 
          \textrm{term}_5 &= -\sqrt{k \tilde{k}} \texttt{t}^{-1}  \, \tr\,R^\nabla_\ww U_+^T \widetilde{R}^\nabla_{\wwb} .  
\end{align*}

The variations of these terms with respect to $g$, using the content of appendix \ref{appendix_EOM}, read
 \begin{align*}
   \delta_g \textrm{term}_1 &= k   \tr \Delta  \left( \nabla_{\ww}U_- R^\nabla_{\wwb} - \nabla_{\wwb }U_+ R^\nabla_{\ww} +\nabla_{\wwb}R^\nabla_{\ww} - [U_+ R^\nabla_{\ww} ,U_- R^\nabla_{\wwb}    ]   \right)  , \\
    \delta_g \text{term}_2 &=-k\,\tr \Delta \Big(\nabla_{\wwb}R^{\nabla}_{\ww}+ F[B]_{\ww\wwb}\Big), \\
  \delta_g \textrm{term}_3 &= \tilde{k}  \tr \Delta [ U_+^T\widetilde{R}^\nabla_{\wwb} , U_-^T\widetilde{R}^\nabla_{\ww}  ]  , \\
   \delta_g \textrm{term}_4 &= \sqrt{\tilde{k} k}  \texttt{t}  \tr \Delta \left(  -\nabla_{\wwb}   U_-^T\widetilde{R}^\nabla_{\ww}  + [U_- R^\nabla_{\wwb} ,   U_-^T\widetilde{R}^\nabla_{\ww} ] \right), \\
    \delta_g \textrm{term}_5  &= -\sqrt{\tilde{k} k} \texttt{t}^{-1} \tr \Delta \left(  -\nabla_{\ww }   U_+^T\widetilde{R}^\nabla_{\wwb}  + [U_+ R^\nabla_{\ww} ,   U_+^T\widetilde{R}^\nabla_{\wwb} ] \right) ,
\end{align*}

where $\Delta \equiv \delta g g^{-1}$. It is then apparent that requiring the sum of the above to vanish for arbitrary $\Delta$ yields the equation of motion \eqref{EOM_g_2d}, after defining the currents \eqref{define_currents}.

\section{More general reductions}\label{general_reductions}
In the treatment of section \ref{section:reduction} we made two key assumptions in the reduction to an $\text{IFT}_2$ each of which can be  relaxed.  In both cases we do not find a NLSM interpretation of the result but are of sufficient interest to warrant a brief comment. 

First, we aligned $\kappa_{\dot{a}} \sim \mu_{\dot{a}}$, i.e.  that the dotted spinor setting reduction to directions matches that involved in setting the boundary conditions.    If this choice is not made a rather elaborate action is obtained, and does not yield a two-dimensionally Lorentz invariant theory in the sense that the Kinetic operator contains not only $\partial_\ww \partial_{\wwb} $ but also $\partial_\ww^2$ and $\partial_{\wwb}^2 $ terms and so the interpretation as a NLSM is limited.  However, there remains some underlying structure since in this more general scenario the Lax operators can be expressed in terms of currents 

\begin{align*}
     J_{\wwb} = \left[  \frac{1}{\aa- \bb } U_- R^\nabla_{\wwb} +  \frac{1}{\aat - \bb } U^T_+ \widetilde{R}^\nabla_{\wwb}  \right]\, , \quad 
     J_{\ww } = \left[  \frac{\aa}{\aa- \bb } U_+ R^\nabla_{\ww} +  \frac{\aat}{\aat - \bb } U_-^T \widetilde{R}^\nabla_{\ww}  \right] \, , \\
         K_{\wwb} = \left[  \frac{1}{\aa- \bb } U_+ R^\nabla_{\wwb} +  \frac{1}{\aat - \bb } U^T_- \widetilde{R}^\nabla_{\wwb}  \right] \, , \quad 
     K_{\ww } = \left[  \frac{\aa}{\aa- \bb } U_- R^\nabla_{\ww} +  \frac{\aat}{\aat - \bb } U_+^T \widetilde{R}^\nabla_{\ww}  \right]  \, , 
\end{align*}
as (n.b. the mixed $\ww$ and $\wwb$ index structure) 
\begin{equation*}
 \begin{aligned}
L^{\kappa}_{\wwb} &= \nabla_{\wwb} + ( b - \zeta) \left( [\kappa \mu] [\hat{\kappa} \hat{\mu}] K_{\wwb} -  [\kappa \hat\mu] [\hat{\kappa}  \mu] J_{\wwb} + [\hat \kappa \mu] [\hat{\kappa} \hat{\mu}]  (   K_{\ww} - J_{\ww})  \right) \,,\\ 
L^{\kappa} _{\ww} &= \nabla_{\ww} +\frac{  ( b - \zeta)}{\zeta}  \left( [\kappa \mu] [\hat{\kappa} \hat{\mu}] K_{\ww} -  [\kappa \hat\mu] [\hat{\kappa}  \mu] J_{\ww} + [ \kappa \mu] [ \kappa \hat{\mu}]  (   K_{\wwb} - J_{\wwb})  \right) \, . 
\end{aligned}
\end{equation*}
In this `non-aligned' scenario we also note that  at the point of enhancement eq.~\eqref{eq:enhancement} under the redefinition $\bar{g}= g \tilde{g}^{-1}$   there is no-longer an elimination of the $g$ degree's of freedom; they are retained in the $K$ currents.  This ultimately due to the fact that the enhanced gauge symmetry in two-dimensions is a result of the residual symmetries preserved by the boundary conditions in four-dimensions.  It is only when the reduction is aligned to boundary conditions that the four-dimensional semi-local symmetries give rise to fully localised gauge freedom in two-dimensions. 

A second consideration in in the reduction ansatz.  In a previous work, \cite{Cole:2024sje}, it was shown that it was possible to relax the ansatz that $B_{\zz}=   B_{\zzb} =0 $.  There choosing $B_{\zz} $ and $B_{\zzb}$ to be constant and in the centre of $\mathfrak{h}$ (but not necessarily central in $\mathfrak{g}$) was shown to modify the resultant two-dimensional gauged WZW by integrable potentials (examples of which include the complex Sine-Gordon model). Restricting  for simplicity again to the aligned $\mu \sim \kappa$ case, we extract the comments   $L_{\ww}$ and $L_{\wwb}$ without imposing any reduction ansatz as 
$$
L_{\wwb} = \nabla_{\wwb} - \pp \nabla_{\zz} + (\bb - \pp) J'_{\wwb} \, , \quad J'_{\wwb} = \left[  \frac{1}{\aa- \bb } ( U_- R^\nabla_{\wwb}-\aa  U_- R^\nabla_{\zz}) +  \frac{1}{\aat - \bb }(  U^T_+ \widetilde{R}^\nabla_{\wwb} -\aat U^T_+ \widetilde{R}^\nabla_{\zz})  \right]\,,
$$
$$
L_{\ww} = \nabla_{\ww} + \frac{1}{\pp} \nabla_{\zzb}+ \pp^{-1} (\bb - \pp)  J'_{\ww } \, , \quad J'_{\ww } = \left[  \frac{1}{\aa- \bb } ( \aa U_+ R^\nabla_{\ww} +  U_+ R_{\zzb})  \frac{1}{\aat - \bb } (\aat U_-^T \widetilde{R}^\nabla_{\ww}  +   U_-^T \widetilde{R}^\nabla_{\zzb} ) \right]\,.
$$
If we choose $\pd_{\zz} = \pd_{\zzb} = 0 $ and also put $B_{\zz} = \phi$, $B_{\zzb} = \bar{\phi}$ as constant and central in $\frak{h}$ we anticipate that the reduced theory will remain integrable and consistent.  In that case we have 
$$
J'_{\wwb} = J_{\wwb} - \bar{{\cal N}}\cdot \phi \, , \quad \bar{{\cal N}}= \frac{\aa }{\aa - \bb} U_- (1- \Ad_g) + \frac{\aat }{\aat - \bb} U_+^T (1- \Ad_{\tilde{g}}) \,,
$$
$$
J'_{\ww} = J_{\ww} +  {\cal N}\cdot \bar{\phi} \, , \quad  {\cal N}= \frac{1 }{\aa - \bb} U_+ (1- \Ad_g) + \frac{1 }{\aat - \bb} U_-^T (1- \Ad_{\tilde{g}}) \,.
$$ 
Focusing our attention at the point of enhancement, eq.~\eqref{eq:enhancement}, and letting  $\phi = \aa^{-1} \Phi$ and $\bar\phi= \aa \bar{\Phi}$ we obtain
 $$
L_{\ww} = \nabla_{\ww} + \frac{1}{ \mathfrak{z}}\bar{\Phi} - \frac{\mathfrak{z}^{-1} + \sigma   }{1 +\sigma }U_+ \left( \bar{R}^\nabla_{\ww}  + \sigma^2 ( 1- \Ad_{\bar{g}})\bar{\Phi}   + (1- \sigma^2) (1- \Ad_g) \bar{\Phi}\right) \,,
$$
  $$
L_{\wwb} = \nabla_{\wwb} - \zeta \Phi   - \frac{ 1 + \zeta   \sigma   }{1 +\sigma }U_-\left( \bar{R}^\nabla_{\wwb}  - \sigma^{-2} ( 1- \Ad_{\bar{g}}) \Phi - (1- \sigma^{-2}) (1- \Ad_g)  \Phi \right) \, .
$$
This is an intriguing situation in which one finds that whilst $g$ no-longer decouples, it appears in the Lax only through the adjoint action on $\Phi$ and $\bar{\Phi}$. If we set $\Phi \propto \bar{\Phi}$ then we see that $g$ enters in the Lax only defined upto the right action of the stabiliser of $\Phi$, which by definition is at least $H$, but could in principle be larger.

\bibliographystyle{JHEP}
\bibliography{main.bib}

\end{document}